\pdfoutput=1
\documentclass[11pt,reqno]{article}
\usepackage[utf8]{inputenc}
\usepackage[T1]{fontenc}
\usepackage{libertine}
\usepackage{libertinust1math}
\usepackage{inconsolata} 
\usepackage{carlito}
\usepackage{microtype}
\usepackage[linesnumbered,noend,ruled]{algorithm2e}
\usepackage{amsmath, amssymb, amsthm}
\usepackage{bm}
\usepackage{graphicx}
\usepackage{thmtools}
\usepackage{thm-restate}
\usepackage{hyperref} 
\usepackage[capitalize, nameinlink]{cleveref}
\usepackage{comment}
\usepackage{enumerate}
\usepackage{enumitem}
\usepackage{float}
\usepackage{fullpage}
\usepackage{mathtools}
\usepackage{subcaption}
\usepackage[dvipsnames]{xcolor}
\usepackage{nag}
\usepackage{authblk}
\usepackage{xfrac}
\usepackage{xspace}
\usepackage{tcolorbox}

\graphicspath{{figures/}{.}}

\newcommand{\declarecolor}[2]{\definecolor{#1}{RGB}{#2}\expandafter\newcommand\csname #1\endcsname[1]{\textcolor{#1}{##1}}}
\declarecolor{White}{255, 255, 255}
\declarecolor{Black}{0, 0, 0}
\declarecolor{LightGray}{216, 216, 216}
\declarecolor{Gray}{127, 127, 127}
\declarecolor{Orange}{237, 125, 49}
\declarecolor{LightOrange}{251,229, 214}
\declarecolor{Yellow}{255, 192, 0}
\declarecolor{LightYellow}{255, 242, 200}
\declarecolor{Blue}{91, 155, 213}
\declarecolor{LightBlue}{222, 235, 247}
\declarecolor{Green}{112, 173, 71}
\declarecolor{LightGreen}{226, 240, 217}
\declarecolor{Navy}{68, 114, 196}
\declarecolor{LightNavy}{218, 227, 243}

\hypersetup{
	colorlinks=true,
	pdfpagemode=UseNone,
	citecolor=Green,
	linkcolor=Navy,
	urlcolor=Navy,
	pdfstartview=FitW,
}

\newlist{longenum}{enumerate}{6}
\setlist[longenum,1]{label=\textbullet,leftmargin=0.40cm}
\setlist[longenum,2]{label=$\circ$}
\setlist[longenum,3]{label=--}
\setlist[longenum,4]{label=\textbullet}
\setlist[longenum,5]{label=$\circ$}
\setlist[longenum,6]{label=--}

\crefformat{equation}{(#2#1#3)}
\crefname{mechanism}{Mechanism}{Mechanisms}

\SetCommentSty{mycommfont}

\DeclareMathOperator*{\argmax}{arg\,max}

\newcommand{\E}{\mathbb{E}}
\newcommand{\R}{\mathbb{R}}

\newcommand{\OPT}{\textnormal{\text{OPT}}\xspace}
\newcommand{\ALG}{\textnormal{\textsf{ALG}}\xspace}

\newcommand{\MaxW}{\textnormal{\textsf{MaxW}}\xspace}
\newcommand{\Gain}{\textnormal{\textsf{Gain}}\xspace}
\newcommand{\AdaptiveGain}{\textnormal{\texttt{adaptive\_gain}}\xspace}
\newcommand{\MarginalGain}{\textnormal{\textsf{MarginalGain}}\xspace}
\newcommand{\Excess}{\textnormal{\texttt{excess}}\xspace}

\newcommand{\Color}{\textnormal{\texttt{active}}\xspace}
\newcommand{\Index}{\textnormal{\texttt{index}}\xspace}
\newcommand{\Mark}{\textnormal{\textsf{Priority}}\xspace}
\newcommand{\Partner}{\textnormal{\texttt{partner}}\xspace}
\newcommand{\blue}{\textnormal{\texttt{true}}\xspace}
\newcommand{\green}{\textnormal{\texttt{false}}\xspace}

\newcommand{\alg}[1]{\textnormal{\textsc{#1}}}
\newcommand{\rv}[1]{\textnormal{\textsf{#1}}} 

\newcommand{\StochasticGreedy}{\alg{StochasticGreedy}\xspace}
\newcommand{\DistributeExcess}{\alg{DistributeExcess}\xspace}
\newcommand{\CaseOne}{lines~13--30\xspace}
\newcommand{\CaseTwo}{lines~31--42\xspace}

\DeclarePairedDelimiter{\set}{\{}{\}}
\DeclarePairedDelimiter{\parens}{(}{)}
\DeclarePairedDelimiter{\bracks}{[}{]}

\theoremstyle{plain}
\newtheorem{theorem}{Theorem}[section]
\newtheorem{lemma}[theorem]{Lemma}

\theoremstyle{definition}
\newtheorem{definition}[theorem]{Definition}

\begin{document}

\title{Online Weighted Matching: Breaking the $\sfrac{1}{2}$ Barrier}

\author[1]{Matthew Fahrbach\thanks{
Email: \href{mailto:matthew.fahrbach@gatech.edu}{\textsf{matthew.fahrbach@gatech.edu}}.
Supported in part by an NSF Graduate Research Fellowship under grant DGE-1650044.
}}
\author[2]{Morteza Zadimaghaddam\thanks{
Email: \href{mailto:zadim@google.com}{\textsf{zadim@google.com}}.
}}
\affil[1]{School of Computer Science, Georgia Institute of Technology}
\affil[2]{Google Z\"urich}

\maketitle

\begin{abstract}

Online matching and its variants are some of the most fundamental
problems in the online algorithms literature.
In this paper, we study the online weighted bipartite matching problem.
Karp~et~al.\ (STOC~1990) gave an~elegant algorithm 
in the unweighted case
that achieves a tight competitive ratio of $1-1/e$.
In the weighted case, however,
we can easily show that no competitive ratio is obtainable
without the commonly accepted free disposal assumption.
Under this assumption, it is not hard to prove that the greedy algorithm
is~$1/2$ competitive, and that this is tight for deterministic algorithms.
We present the first randomized algorithm that breaks this long-standing $1/2$ barrier
and achieves a competitive ratio of at least $0.501$.
In light of the hardness result of Kapralov et al.\ (SODA 2013) that restricts
beating a $1/2$ competitive ratio for the monotone submodular welfare maximization
problem, our result can be seen as strong evidence that solving the
weighted bipartite matching problem is strictly easier than submodular welfare
maximization in the online setting.
Our approach relies on a very controlled use of randomness, which allows our
algorithm to safely make adaptive decisions based on its previous assignments.

\end{abstract}

\pagenumbering{gobble}
\clearpage
\pagenumbering{arabic}

\newpage
\tableofcontents
\pagenumbering{gobble}
\clearpage
\pagenumbering{arabic}

\section{Introduction}
\label{sec:introduction}

Matchings are fundamental structures in graph theory 
that play a crucial role in combinatorial optimization.
An enormous amount of effort has gone into designing efficient
algorithms for finding maximum matchings
in terms of cardinality or the total weight of its allocation.
In particular, matchings in bipartite graphs have found countless applications
in settings where it is desirable to assign entities from one set to
those in another set (e.g., matching students to schools,
physicians to hospitals,
computing tasks to servers,
and impressions in online media to advertisers).
Due to the tremendous growth of matching markets in digital domains,
efficient online matching algorithms have become increasingly important.
In particular, search engine companies have created opportunities for online
matching algorithms to have enormous impact in multi-billion dollar advertising markets.
Motivated by these applications, we consider the problem of matching a set~$I$ of
impressions that arrive one by one to a set $A$ of advertisers that are given
in advance.
When an impression arrives, its edges to the advertisers are revealed
and an irrevocable decision has to be made about which advertiser the impression
should be assigned to.
Karp et al.~\cite{karande2011online}
gave an elegant online algorithm called \textsc{Ranking} to find matchings
in unweighted bipartite graphs with a \emph{competitive ratio} of $1-1/e$,
and they also proved that this is the best competitive ratio that can be achieved.

The situation in the weighted case is much less clear.
This is partly due to the fact that no competitive algorithm can be designed
without the \emph{free disposal assumption}. To see this, consider two
instances of the online bipartite weighted matching problem, each with one advertiser $a$ and two impressions.
The weight of the first impression to $a$ is $1$ in
both instances, and the weight of the second impression to $a$ is zero in the first instance
and $L$, for some arbitrarily large $L$, in the second instance.
The online algorithm has no way to distinguish between these two instances,
even after the first impression arrives.
When the first impression arrives,
the algorithm must decide whether or not to assign this impression to $a$.
Not assigning it gives a competitive ratio of zero for the first instance,
and assigning it gives a competitive ratio of $1/L$, which can be arbitrarily
small, for the second.
Note that assigning both impressions to $a$ is not an option in this setting.
This example is the primary motivation for allowing assignments of multiple
impressions to a single advertiser.

\textbf{Free Disposal Assumption.}
In display advertising applications,
assigning more impressions to an advertiser than they paid for
only makes them happier.
In other words, we can assign more than one impression to any
given advertiser $a \in A$. However, instead of achieving the weights of all
edges assigned to~$a$,
we only receive the maximum weight of the edges assigned to $a$.
Concretely, the total weighted achieved by an allocation is equal
to $\sum_{a \in A} \max_{i \in \rv{T}_a} w_{i,a}$, where $\rv{T}_a$ is the set
of impressions assigned to $a$ and~$w_{i,a}$ is the weight of the edge between
$i$ and $a$.
In the display advertising literature~\cite{feldman2009online,korula2013bicriteria}, 
the free disposal assumption
is well received and widely applied because of its natural economic interpretation.
With free disposal, it is not hard to
reduce weighted bipartite matching to
the monotone submodular welfare maximization problem, where we can
apply known $1/2$-competitive greedy algorithms~\cite{fisher1978analysis,lehmann2006combinatorial}.

\subsection{Our Contributions}

For almost thirty years since the seminal work of
Karp et al.~\cite{karp1990optimal},
finding an algorithm for the online weighted bipartite matching problem
that achieves a competitive ratio greater than $1/2$
has been a tantalizing open problem.
In this paper, we introduce the $\StochasticGreedy$ algorithm and answer this question in the affirmative,
breaking the long-standing $1/2$ competitive ratio barrier
(under the free disposal assumption).

\begin{theorem}
There exists a 0.501-competitive algorithm for the
online weighted bipartite matching problem.
\end{theorem}

\noindent
Given the hardness result of Kapralov et al.~\cite{kapralov2013online} that
restricts beating a competitive ratio of $1/2$ for monotone submodular welfare
maximization, our algorithm can be seen as strong evidence that solving
weighted bipartite matching is strictly easier than submodular welfare maximization
in an online setting.

One of our main technical contributions
is the controlled use of randomness in \StochasticGreedy,
which allows the algorithm to safely make adaptive
decisions based on past assignments
and prevents the cascading of conditional probabilities in our analysis.
A more subtle feature of our use of randomness is that in every step of the
algorithm, expected values of random variables are
computed over all possible branches of the randomized algorithm instead of
being conditioned on the current state.
This ensures that most sequences of variables in the algorithm are deterministic
quantities governed solely by the input graph and arrival order of the
impressions.
Our method for making adaptive decisions combined with the
limited randomness of the algorithm allows us to efficiently maintain
these expected values 
(over all possible branches of the algorithm) using dynamic programming.
Lastly, we introduce a mechanism called \DistributeExcess
in our analysis, which is not part of the algorithm
but allows us to systematically redistribute the extra marginal gain that the
algorithm produces and improve upon the greedy algorithm.

\subsection{Related Works}

We first draw attention to two very recent works of Huang and Tao~\cite{huang2019unweighted,huang2019weighted}
that build on an earlier version of this paper that appeared on arXiv in 2017~\cite{zadimoghaddam2017online}.
These works ``distill a key ingredient from the algorithm of Zadimoghaddam''
and enhance this idea by using the online primal-dual framework~\cite{devanur2013randomized}
and an improved \emph{online correlated selection} scheme to achieve an
improved competitive ratio of $0.514$ for the same online weighted bipartite matching problem.
In this version of our paper, we have tried to improve and simplify the
presentation of our algorithm and analysis
so that our approach is easier to understand.

The literature online weighted bipartite matching algorithms is extensive,
but most of these works are devoted to achieving competitive ratios greater than $1/2$
(usually $1-1/e$ or $1-\varepsilon$)
by assuming that advertisers have large capacities or that some
stochastic information about the arrival order of the impressions is known in advance.
There are many exciting papers in this area, but we only list a few of the
leading works and refer interested readers to the excellent survey
of Mehta~\cite{mehta2013online}.
We note that there have recently been several significant advances in more general
settings, including different arrival models
and general (non-bipartite) graphs~\cite{huang2018match,gamlath2019beating,gamlath2019online,huang2019tight}. 

\textbf{Large Capacities.}
Exploiting the large capacities assumption to beat the $1/2$ competitive ratio
barrier dates back two decades ago to~\cite{kalyanasundaram2000optimal}.
Feldman et al.~\cite{feldman2009online} gave a primal-dual algorithm
that achieves a competitive ratio of $1-1/e$ assuming that each advertiser
has a large capacity, where the capacity denotes the number of
impressions that can be assigned to it (e.g., the Display Ads problem).
Under similar assumptions, the same competitive ratio was obtained for the
Budgeted Allocation problem~\cite{mehta2005adwords, buchbinder2007online},
in which advertisers have some budget constraint on the total weight that can
be assigned to them rather than the number of impressions.
From a theoretical point of view, one of the primary goals in the online
matching literature is to provide algorithms with competitive ratio
greater than $1/2$ without making any assumption on the capacities of advertisers.
Without loss of generality, we can assume every advertiser
has capacity~one, since we can replace each advertiser $a$ 
with capacity $C_a$ by $C_a$ identical copies of $a$, each with unit capacity.

\textbf{Stochastic Arrivals.}
If we have knowledge about the arrival patterns of impressions,
then we can often leverage this information to design better algorithms.
Typical stochastic assumptions include
assuming the impressions are drawn from some known or unknown
distribution~\cite{feldman2009online2,karande2011online,devanur2011near,haeupler2011online,
manshadi2012online,
mehta2012online,jaillet2013online}
or that the impressions arrive in a random
order~\cite{goel2008online,devanur2009adwords,feldman2010online, mahdian2011online,
mirrokni2012simultaneous,mehta2015online,huang2019online}.
These works achieve a $1-\varepsilon$ competitive ratio if the large
capacity assumption holds in addition to the stochastic assumptions,
or at least $1-1/e$ for arbitrary capacities.
Korula et al.~\cite{korula2018online} show that the greedy algorithm
is $0.505$-competitive for the more general problem of submodular welfare
maximization if the impressions arrive in random
order, without making any assumptions on the capacities.
The random order assumption is particularly justified because
Kapralov et al.~\cite{kapralov2013online} show that beating $1/2$ for
submodular welfare maximization in the oblivious adversary model
is equivalent to proving $\textbf{NP} = \textbf{RP}$.


\section{Preliminaries}
\label{sec:preliminaries}


Let $A$ be the set of advertisers,
$I$ be the set of $n$ impressions, and
$w_{i,a}$ denote the nonnegative weight of the edge between impression $i$
and advertiser $a$.
If there is not an edge between $i$ and $a$, we introduce an edge
of weight zero to simplify the notation.
The set of advertisers is given in advance, and
the impressions arrive one by one 
at times $t=1,2,\dots,n$.
We do not assume $n$ is known to the algorithm.
When an impression~$i$ arrives at time $t_i$, 
all edge weights incident to $i$ are revealed to the algorithm, that is, $w_{i,a}$ for all $a \in A$,
and the algorithm has to assign $i$ to one of the advertisers at this time.
This is an irrevocable decision and cannot be changed later.
At the end of the algorithm, if more than one impression is assigned to an
advertiser, only the impression with the maximum weight is kept.
The rest are discarded and do not counted towards the total weight of the allocation.
This is known as the \emph{free disposal assumption}.
The objective is to maximize the total weight of maximum-valued edges
assigned to the advertisers, that is, 
$\sum_{a \in A} \max_{i \in \rv{T}_a} w_{i,a}$,
where $\rv{T}_a$ is the set of impressions assigned to $a$.

In this paper, we assume that the impressions arrive in an adversarial order.
Specifically, we deal with an \emph{oblivious adversary}, that is,
an adversary that does not have access to the outcomes of the random choices made by the algorithm,
and instead has to fix the input graph and arrival order 
in advance.
We use the standard notion of \emph{competitive ratio} to measure the
performance of our online algorithm.
For a randomized bipartite matching algorithm in this adversarial model,
the competitive ratio is defined to be
the worst-case ratio $\min_{G(A,I,w), \text{ order of $I$}} \E[\ALG]/\OPT$,
where $\ALG$ is a random variable denoting the value of the objective
function achieved by the algorithm
and
$\OPT$ is the maximum objective value attained offline.

We present our main randomized online algorithm \StochasticGreedy in \Cref{sec:algorithm}.
This algorithm uses randomness in a very controlled manner, so 
we deliberately use a sans serif font and ``upper camel case'' to
distinguish quantities that are random variables.
To evaluate how much marginal value (i.e., increase in the objective function)
can be achieved by assigning an
impression $i$ to an advertiser $a$ at every point in the algorithm,
we need to keep track of the maximum weight assigned to $a$ by \StochasticGreedy.
Therefore, we let the random variable $\MaxW_{a}^t$
denote the maximum weight assigned to $a$ by \StochasticGreedy
up to (and including) time $t$ for every $0 \le t \le n$.
Since assignments are made at times $t=1,2,\dots,n$, we define $\MaxW_a^0$
to be zero for all $a \in A$.
Next, we define the random variable $\Gain_{i,a}$ to be the marginal gain
of assigning impression $i$ to advertiser~$a$.
Formally, we have $\Gain_{i,a} = (w_{i,a} - \MaxW_{a}^{t_i - 1})^+$, where
$(x)^+$ is the function $\max\{0,x\}$
and 
$t_i$ is the arrival time of impression $i$.
We note that $\Gain_{i,a}$ depends on the random choices that
\StochasticGreedy makes before~$i$ arrives.

We let $\ALG$ be the total weight achieved by \StochasticGreedy.
Since only the maximum weight edge assigned to each advertiser contributes to
the total weight of the final allocation, we have
$\ALG = \sum_{a \in A} \MaxW_{a}^n$.
We can also interpret the total weight by letting
$\ALG = \sum_{i \in I} \MarginalGain_i$, where $\MarginalGain_i$
is the random variable that denotes how much the assignment of impression $i$
increases the total weight of the allocation at the time of its assignment.
We let $\OPT$ denote the maximum weight matching of the instance, and we let
$a_i^*$ be the advertiser to which impression $i$ is assigned in $\OPT$.
We overload the notation of $\OPT$ and also let it be the weight
of the allocation, that is, $\OPT = \sum_{i \in I} w_{i,a_{i}^*}$.
For the sake of analysis, we can add a large enough number of dummy
impressions (advertisers) with edges of weight zero to all advertisers (impressions)
so that all vertices (impressions and advertisers) are matched in the optimal solution.
For the completeness of our algorithm, we also introduce a null impression
$i=0$ with weight $w_{0,a} = 0$ for all $a \in A$.

\section{The \StochasticGreedy Algorithm}
\label{sec:algorithm}

In this section we introduce our randomized online algorithm \StochasticGreedy
(\Cref{alg:stochastic_greedy}).
We start by describing the algorithm at a high level,
and then we present it formally together with two important
lemmas that highlight its deterministic features.
In \Cref{subsec:variables}
we describe the variables in the algorithm.
Then in \Cref{subsec:case_one} and \Cref{subsec:case_two} we discuss the
two main cases the algorithm considers when assigning an impression.
Lastly, in \Cref{subsec:intuition} we explain why
this approach breaks the $1/2$ competitive ratio barrier.

Our algorithm builds on the greedy approach.
Upon the arrival of impression $i$, \StochasticGreedy first constructs
a set $B \subseteq A$ of candidate advertisers that can potentially be matched with $i$.
If there are multiple candidates, the algorithm considers the top two $a_1$ and $a_2$,
and performs one of the three actions uniformly at random:
(1) greedily assign $i$ to $a_1$,
(2) greedily assign $i$ to $a_2$,
or (3) \emph{adaptively} choose between $a_1$ and $a_2$ by
looking at a past decision of the algorithm.
The top candidates are determined
by their expected gain $\E[\Gain_{i,a}]$
(where the randomness is over all possible branches of the algorithm up to this point and
not conditioned on the current state)
and an \emph{adaptive gain} value that
originates in the proof of~\Cref{lem:marginal_gain_lower_bound}.
In the event that there are not multiple candidates,
the algorithm makes a nonadaptive assignment.

To adaptively decide between the top two candidates,
the algorithm looks
at the result of the last coin flip associated with $a_k$, where $a_k$ is the
advertiser in $\{a_1,a_2\}$ with the greater adaptive gain.
If the assignment based on this coin flip was adaptive,
then the algorithm chooses between $a_1$ and~$a_2$ uniformly at random.
Otherwise, the assignment of the past impression $i'$ associated with this coin flip
was nonadaptive, and the algorithm looks at whether or not $i'$ was matched to $a_k$.
If $i'$ was matched to~$a_k$ then the algorithm assigns $i$ to the other
top candidate in $\{a_1,a_2\}$,
and if $i'$ was not matched to $a_k$ then $i$ is assigned to $a_k$.
In general, adaptive decisions can cause cascading effects of conditional
probabilities that can severely alter the distribution of many $\MaxW_{a}^t$ variables.
However, by ensuring that the adaptive decisions are based on an earlier
nonadaptive (i.e., random) assignment,
we can prevent this effect and analyze the algorithm.
We continue this discussion about the benefits of this kind of adaptive decision
in more detail in \Cref{subsec:intuition}.

Now we formally present \StochasticGreedy in \Cref{alg:stochastic_greedy}.
This algorithm takes as input two nonnegative parameters $\varepsilon$ and $\delta$
that control the thresholds for how greedy and adaptive the algorithm is, respectively.
We optimize these constants later in our analysis in \Cref{sec:analysis}.
While \Cref{alg:stochastic_greedy} is initially cumbersome~to~parse,
we point out that it is comprised of two separate cases that can be analyzed individually
(see \Cref{subsec:case_one} and \Cref{subsec:case_two}).
We also acknowledge that the definitions of the $\AdaptiveGain_{i,a}$ variables
and the set~$B$ initially appear to be unnatural,
but these conditions arise in our analysis
and allow us to overcome the shortcomings of the greedy algorithm.
Before stepping through the details of \Cref{alg:stochastic_greedy}, 
we first make two critical observations about how the algorithm uses randomness.

\begin{restatable}[]{lemma}{Deterministic}
\label{lem:deterministic}
The only random variables in $\StochasticGreedy$ are the assignments of the
impressions and the values of $\Mark_{a}$ and $\rv{R}_i$.
All other variables (e.g., the maximum gains $M_i$,
  all values of $\AdaptiveGain_{i,a}$,
  the sequence of sets $B$ and choices $a_1$ and $a_2$,
  all updates to $\Color(a), \Index(a)$, $\Partner(a)$, and
  the sums $S_a$)
are deterministic quantities that depend solely on the
instance and the arrival order of the impressions.
\end{restatable}

\begin{restatable}[]{lemma}{ExpectedGainValues}
\label{lem:expected_gain_values}
We can maintain the probability mass function for all random variables
$\Gain_{i,a}$ over the course of algorithm.
In particular, we can efficiently compute the value $\E[\Gain_{i,a}]$
at any point in \StochasticGreedy.
\end{restatable}

\noindent
In particular,~\Cref{lem:deterministic} guarantees that the top candidates in each
step are predetermined by the input instance,
even though the assignments of past impressions to these advertisers
could have been random.
This property is simple to show by induction but easy to miss because of the
complexity of~\Cref{alg:stochastic_greedy}.
\Cref{lem:expected_gain_values}
states that we can efficiently compute $\E[\Gain_{i,a}]$ for all $a \in A$
in line~5 of \Cref{alg:stochastic_greedy}. 
This is a consequence of the limited randomness in the algorithm
and dynamic programming.
We defer the proofs of both of these lemmas to~\Cref{app:algorithm}.

\begin{algorithm}
\caption{Online weighted bipartite matching algorithm.}
\label{alg:stochastic_greedy}
\DontPrintSemicolon

\SetKwProg{FunctionStochasticGreedy}{\textbf{function} $\StochasticGreedy(\varepsilon,\delta)$}{}{}

\FunctionStochasticGreedy{}{
    Set $\Color(a) \gets \green$,
        $\Index(a) \gets 0$,
        $\Partner(a)\gets 0$,
        $\Mark_a \gets 0$,
        $S_a \gets 0$
        for all $a \in A$\;
    \For{$t=1,2,\dots,|I|$}{
      Let $i$ be the impression that arrives at time $t$, i.e., $t_i = t$\; 
      $M_i \gets \max_{a \in A} \E[\Gain_{i,a}]$
        \tcp*[r]{Not conditioned on the algorithm's state (see \Cref{lem:deterministic})}
      \For(\tcp*[f]{Compute adaptive gain values}){$a \in A$}{
        \If{$\Color(a) = \blue$ \textnormal{and} $w_{i,a} \ge w_{\Index(a),a} - \delta M_i$}{
          $\AdaptiveGain_{i,a} \gets
            (\E[\Gain_{\Index(a),a}]/3 - (w_{\Index(a),a} - w_{i,a})^+/3 - S_a)^+/12$\;
        }
        \Else{$\AdaptiveGain_{i,a} \gets 0$}
      }
      $B \gets \set{a \in A : (w_{i,a} \ge w_{\Index(a),a} - \delta M_i)
        \text{ and }
        (\E[\Gain_{i,a}] + 2/3 \cdot \AdaptiveGain_{i,a} \ge (1-\varepsilon)M_i)
      }$\;
      Let $\rv{R}_i$ be a uniformly random real number in the interval $[0,1)$\;
      \If(\tcp*[f]{Case I: There are enough candidates to exploit adaptivity}){$|B| \ge 2$}{
        $a_1 \gets \argmax_{a \in B} \E[\Gain_{i,a}] + 2/3 \cdot \AdaptiveGain_{i,a}$\;
        $a_2 \gets \argmax_{a \in B\setminus\{a_1\}} \E[\Gain_{i,a}] + 2/3 \cdot \AdaptiveGain_{i,a}$\;
        \For(\tcp*[f]{Couple the top pair of advertisers $a_1$ and $a_2$}){$a \in \{a_1,a_2\}$}{
          Set
          $\Color(a) \gets \blue$, 
          $S_a \gets 0$, 
          $\Index(a) \gets i$\;
          \If{$\Partner(a) \not\in \{a_1,a_2\}$}{$\Color(\Partner(a)) \gets \green$}
        }
        Set $\Partner(a_1) \gets a_2$ and $\Partner(a_2) \gets a_1$\;
        $k \gets \argmax_{j \in \{1,2\}} \AdaptiveGain_{i,a_j}$\;
        \If(\tcp*[f]{Make an adaptive decision if possible}){$\rv{R}_i \in [0,\sfrac{1}{3})$ \textnormal{or} $\AdaptiveGain_{i,a_k} = 0$}{
          \lIf{$\Mark_{a_k} = 2$ \textnormal{and} $\AdaptiveGain_{i,a_k} > 0$}{Assign $i$ to $a_{k}$}
          \lIf{$\Mark_{a_k} = 1$ \textnormal{and} $\AdaptiveGain_{i,a_k} > 0$}{Assign $i$ to $\Partner(a_k)$}
          \If{$\Mark_{a_k} = 0$ \textnormal{or} $\AdaptiveGain_{i,a_k} = 0$}{
            Assign $i$ to $a_1$ or $a_2$ each with probability $\sfrac{1}{2}$\;
          }
          Set $\Mark_{a_1} \gets 0$ and $\Mark_{a_2} \gets 0$\;
        }
        \Else(\tcp*[f]{Make a random assignment and prepare for future adaptive decisions}){
          \lIf{$\rv{R}_i \in [\sfrac{1}{3},\sfrac{2}{3})$}{Assign $i$ to $a_1$ and set $\Mark_{a_1} \gets 1$, $\Mark_{a_2} \gets 2$}
          \lIf{$\rv{R}_i \in [\sfrac{2}{3},1)\hspace{0.10cm}$}{Assign $i$ to $a_2$ and set $\Mark_{a_1} \gets 2$, $\Mark_{a_2} \gets 1$}
        }
      }
      \Else(\tcp*[f]{Case II: There is no adaptivity}){
        $B' \gets \{a \in A : (w_{i,a} \ge w_{\Index(a),a} - \delta M_i)
        \textnormal{ and }
        (\E[\Gain_{i,a}] \ge (1-\varepsilon)M_i)
        \}$
        \tcp*[r]{Note $B' \subseteq B$}
        $C \hspace{0.078cm}\gets \{a \in A : (w_{i,a} < w_{\Index(a),a} - \delta M_i)
        \textnormal{ and }
        (\E[\Gain_{i,a}] \ge (1-\varepsilon)M_i)
        \}$\;
        \If(\tcp*[f]{Advertiser $a_2$ does not exist}){$|B' \cup C| = 1$}{
          Assign $i$ to $a_1 \gets \argmax_{a \in A} \E[\Gain_{i,a}]$\;
          Set $S_{a_1} \gets S_{a_1} + M_i$\;
        }
        \Else(\tcp*[f]{Make a random assignment to the top two choices}){
          \lIf{$B' \ne \emptyset$}{$a_1 \gets$ the only advertiser in $B'$}
          \lElse{$a_1 \gets \argmax_{a \in C}\E[\Gain_{i,a}]$}
          $a_2 \gets \argmax_{a \in C\setminus\{a_1\}} \E[\Gain_{i,a}]$\;
          Assign $i$ to $a_1$ or $a_2$ each with probability $\sfrac{1}{2}$\;
          Set $S_{a_1} \gets S_{a_1} + M_i/2$ and
          $S_{a_2} \gets S_{a_2} + M_i/2$\;
        }
      }
    }
}
\end{algorithm}

We begin the description of \StochasticGreedy by explaining
the preprocessing stage in lines~5--11 of \Cref{alg:stochastic_greedy}.
When impression $i$ arrives, the algorithm first computes
the maximum expected marginal gain $M_i = \max_{a \in A} \E[\Gain_{i,a}]$
as a benchmark.
We remark that it is not hard to show that the standard $1/2$ competitive ratio
proof of the greedy algorithm goes through if we use $\E[\Gain_{i,a}]$ instead
of their realized values.
Using expected values, however, has the advantage that if there are two
choices with high expected gains, then the algorithm can
occasionally realize them in a controlled way and exploit the gap between
them to achieve a better result.
Next, for every $a \in A$ the algorithm computes their adaptive gain value,
and then it constructs the set of candidates $B$.
There is some slack in how greedy \Cref{alg:stochastic_greedy} is,
but an advertiser must be able to contribute a value of
at least $(1-\varepsilon)M_i$ to be considered.
We explain the meaning of the variables used in lines~7--11 in the next subsection,
but for now we note that all of the quantities in these formulas are deterministic
and well-defined.
If there are at least two candidates in $B$, the algorithm assigns~$i$
in Case~I (\CaseOne). Otherwise, the algorithm jumps to line~31 and assigns~$i$
in Case~II (\CaseTwo).
We explain these cases in \Cref{subsec:case_one} and \Cref{subsec:case_two},~respectively.

\subsection{Variable Descriptions}
\label{subsec:variables}
All of the following variables with the exception of $\AdaptiveGain_{i,a}$
help maintain the state of \Cref{alg:stochastic_greedy}.
We reiterate that $\Mark_a$ and $\rv{R}_i$ are the only variables that are randomized.
All other variables (at every time step of the algorithm) are predetermined
by the input graph and the arrival order of the impressions.

\begin{itemize}
  \itemsep0em 
  \item $\Color(a)$ is a boolean value that indicates whether or not advertiser $a$
    potentially has an adaptive gain.
    At its core,
    this variable serves as a mechanism for checking if $a$ is 
    a partner in a valid pairing.

  \item $\AdaptiveGain_{i,a}$
    represents how much extra marginal gain that algorithm can achieve
    by adaptively assigning $i$ to $a$.
    If an adaptive assignment is made, the algorithm achieves an additional
    marginal gain
    that is at most proportional to $\E[\Gain_{\Index(a),a}]$
    since it knows the earlier impression $\Index(a)$ was nonadaptively
    assigned to $\Partner(a)$ and not $a$.
    We give further intuition for this mechanic in~\Cref{subsec:intuition}
    and the derivation of this formula in
    the proof of \Cref{lem:marginal_gain_lower_bound}.

  \item $\Index(a)$ records the last impression for which $a$ was considered
    in a potentially adaptive assignment.
    If $i$ is to be matched to $a_1$ or $a_2$ in \CaseOne,
    the algorithm sets $\Index(a_1) \gets i$ and $\Index(a_2) \gets i$.

  \item $\Partner(a)$
    records the last advertiser with which $a$ was paired in a potentially
    adaptive assignment.
    If $i$ is to be matched to $a_1$ or $a_2$ in \CaseOne,
    then we set $\Partner(a_1) \gets a_2$ and $\Partner(a_2) \gets a_1$.

  \item $\Mark_a$ is a random variable for the outcome of the last potentially
    adaptive assignment involving~$a$.
    If $i$ is to be matched to $a_1$ or $a_2$ in \CaseOne,
    one of the following actions is performed uniformly at random:
    (1) greedily assign $i$ to $a_1$ and set $\Mark_{a_1} \gets 1$, $\Mark_{a_2} \gets 2$;
    (2) greedily assign $i$ to $a_2$ and set $\Mark_{a_1} \gets 2$, $\Mark_{a_2} \gets 1$;
 or (3) adaptively assign $i$ to $a_1$ or $a_2$ by looking
    at $\Mark_{a_k}$, where $k \in \{1,2\}$ is defined in line~21,
    and reset $\Mark_{a_1} \gets 0, \Mark_{a_2} \gets 0$.
    The intuition here is that in the first two actions,
    the advertiser that is not matched with $i$ receives a higher
    priority value and is therefore more likely to be adaptively assigned
    a future impression.
    In the third action, an adaptive decision is potentially made based
    on these priority values, and then the priorities are reset.

  \item $S_a$ is an upper bound for the sum of expected gains 
    assigned to $a$ in \CaseTwo
    since the last time~$a$ was chosen as a candidate in \CaseOne.
    Whenever $a$ is a choice in \CaseOne, $S_a$ is reset to zero.
\end{itemize}


\subsection{Case I: Lines~13--30 and Adaptive Assignments}
\label{subsec:case_one}

If $B$ contains at least two advertisers,
then $a_1$ and $a_2$ are chosen as the top two candidates in lines~14--15
based on their value of
$\E[\Gain_{i,a}] + 2/3 \cdot \AdaptiveGain_{i,a}$.
Before assigning $i$ in lines~21--30, the algorithm performs
an update procedure in lines~16--20 to couple the pair of advertisers $a_1$
and $a_2$.
For each of the top candidates $a \in \{a_1,a_2\}$,
this step
activates $a$ for future a adaptive decision,
deactivates its previous partner (unless this partner is the other top candidate),
and updates $\Partner(a)$ to be the other top candidate.
This routine ensures that the $\Color$ variables are set to $\texttt{true}$
if and only if their advertiser is in a proper pairing.
This is important because the algorithm uses the $\Color$ state of an
advertiser to ensure that $\AdaptiveGain_{i,a}$
is set to zero for all unpaired advertisers $a$ in lines~7--10.
Finally, for each $a \in \{a_1,a_2\}$, line~17
also updates $\Index(a)$ to the current impression and resets the value of $S_a$.

Now we focus on the assignment of $i$ in lines~21--30.
The algorithm starts by drawing a uniformly random real number $\rv{R}_i$ from
the interval $[0,1)$.
We start by describing the assignment rules in lines~29--30, as they are simpler
to explain and give insight into how the adaptive decision works.
With probability~$1/3$ (i.e., the event $\rv{R}_i \in [\sfrac{1}{3}, \sfrac{2}{3})$ in line~29),
the algorithm matches $i$ to $a_1$.
We note that this assignment
does not depend on any previous coin flips of the algorithm because the
sequence of top candidates over the course of the algorithm
is determined by the input instance (\Cref{lem:deterministic}).
It will be useful for future adaptive assignments to record that the
algorithm nonadaptively assigned $i$ to $a_1$,
so we update $\Mark_{a_1} \gets 1$ and $\Mark_{a_2} \gets 2$.
Formally, $\Mark_{a} = 1$ means that the last time $a$ was chosen as a
candidate in \CaseOne, the impression that arrived at this time was nonadaptively
assigned to $a$.
The event $\Mark_{a} = 2$ is in some sense the complement and means that the
last time $a$ was chosen as a candidate in \CaseOne, the impression that arrived
was nonadaptively assigned to the other top candidate (i.e., not $a$).
We refer to this variable for the adaptive state of an advertiser as a priority because if in \CaseOne a
nonadaptive decision is made and the top candidate does not receive the impression,
then it is given a higher priority and is more likely to 
be assigned an impression in the future.
With another probability of $1/3$
(i.e., when $\rv{R}_i \in [\sfrac{2}{3}, 1)$ in line~30),
the algorithm assigns $i$ to $a_2$ and updates the $\Mark$ variables accordingly.

With the remaining $1/3$ probability
(i.e., when $\rv{R}_i \in [0,\sfrac{1}{3})$ in line~22),
impression $i$ is assigned adaptively.
We first note that after this assignment,
$\Mark_{a_1}$ and $\Mark_{a_2}$ are reset to $0$.
To simplify the description of this part of the algorithm, assume that
$\AdaptiveGain_{i, a_1} \ge \AdaptiveGain_{i, a_2}$.
This means $k$ is set to~$1$ in line~21.
The assignment of $i$ is conditioned on the
assignment of $i'$, where $i'$ is the last impression
that chose $a_1$ as a candidate in \CaseOne.
Note that $i'$ was the value of $\Index(a_1)$ immediately before its update in line~17.
The algorithm adaptively makes its assignment conditioned on past events
by looking at the value of $\Mark_{a_1}$.
If $i'$ was matched nonadaptively to $a_1$, then the algorithm makes an adaptive choice
and assigns $i$ to $a_2$ (i.e., the case where $\Mark_{a_1} = 1$ and $\Mark_{a_2} = 2$).
Intuitively, this is beneficial because
conditioning on the event where $i'$ was assigned to $a_1$ decreases the expected
gain of assigning~$i$ to $a_1$. Thus, we should consider $a_2$ as the better option.
Similarly, if $\Mark_{a_1} = 2$ then the algorithm adaptively assigns $i$ to $a_1$.
The last case is when $\Mark_{a_1} = 0$.
Since we want to prevent the chaining of conditional probabilities,
the algorithm does not make an adaptive choice here and instead assigns $i$ to $a_1$ or~$a_2$
uniformly at random using a new coin toss that is independent of $\rv{R}_i$.
In the event that $\AdaptiveGain_{i, a_1} = 0$
(which means that $\AdaptiveGain_{i, a_2} = 0$ because
we assumed $\AdaptiveGain_{i,a_1} \ge \AdaptiveGain_{i,a_2}$),
it suffices for the algorithm to make a random assignment in line~26.

\subsection{Case II: Lines~31--42}
\label{subsec:case_two}

If the set $B$ contains zero or one advertiser,
then the algorithm is forced to make a nonadaptive assignment.
The high level idea in this case is that we want to choose at most two
advertisers to be matched with~$i$
while ensuring that both of the following conditions are met:
\begin{itemize}
  \itemsep0em 
  \item The advertiser with maximum expected gain, $\argmax_{a \in A} \E[\Gain_{i,a}]$,
    is chosen as one of the candidates.
  \item If $B$ is not empty and the only advertiser in $B$ has expected
    gain at least $(1-\varepsilon)M_i$, it should be chosen.
\end{itemize}

First observe that the definitions of the sets $B'$ and $C$
in lines~32--33 imply that their union $B' \cup C$ consists of all advertisers with
expected gain at least $(1-\varepsilon)M_i$.
Therefore, if the conditional statement on line~34 holds, the advertiser
$\argmax_{a \in A} \E[\Gain_{i,a}]$ is the only advertiser that meets
the $(1-\varepsilon)M_i$ threshold.
The algorithm assigns $i$ to this advertiser and does not consider a second option.
Otherwise, the algorithm selects the only advertiser in $B' \subseteq B$ (if it exists)
as a candidate
and chooses one or two additional advertisers in~$C$ with the highest expected gains
to be the candidates $a_1$ and $a_2$.
The impression $i$ is then assigned to $a_1$ or $a_2$ with equal probability.
The only remaining detail is the variable $S_a$, which maintains an upper bound
for the sum of expected gains assigned to $a$ since the last time $a$ was
chosen as a candidate in \CaseTwo.
We note that $S_{a}$ is reset to zero in line~17,
and is otherwise incremented in line~36 or line~42 in a way that is consistent
with the probability of impression $i$ being assigned to $a$.


\subsection{Intuition for Breaking the $\sfrac{1}{2}$ Barrier}
\label{subsec:intuition}

The key result that gives us a chance to break the $1/2$ barrier
is \Cref{lem:marginal_gain_lower_bound}, which states that all assignments in
\CaseOne 
satisfy
$\E[\MarginalGain_i] \ge (\E[\Gain_{i,a_1}] + \E[\Gain_{i,a_2}])/2 + \AdaptiveGain_{i, a_1} + \AdaptiveGain_{i, a_2}$.
Recall from line~7 of \Cref{alg:stochastic_greedy} that
$\AdaptiveGain_{i,a} = (\E[\Gain_{\Index(a),a}]/3 - (w_{\Index(a),a} - w_{i,a})^+/3 - S_a)^+/12$.
We are able to prove \Cref{lem:marginal_gain_lower_bound} because we ensure
that adaptive decisions are based on a past nonadaptive assignment, which stops
the chaining of conditional dependencies.
We first argue that $\AdaptiveGain_{i,a}$ can be thought of
as some constant fraction of $\E[\Gain_{\Index(a),a}]$.
If this is not the case, then $(w_{\Index(a),a} - w_{i,a})^+$ or~$S_a$ is large.
The condition $w_{i,a} \ge w_{\Index(a),a} - \delta M_i$ on line~7 implies
that $(w_{\Index(a),a} - w_{i,a})^+ \le \delta M_i$ is not too large,
so we can bound the drop in $\E[\Gain_{\Index(a),a}]$.
If $S_a$ is large, then the algorithm made potential assignments to~$a$ in
\CaseTwo.
Assignments to $a$ in \CaseTwo are favorable because
they either agree with $\OPT$ or yield substantially more gain than
$\E[\Gain_{i,a_i^*}]$.
Ultimately, this allows us to charge the additional gain from these assignments
to $\AdaptiveGain_{i,a}$.

We proceed by assuming that every $\AdaptiveGain_{i,a}$ variable
is proportional to $\E[\Gain_{\Index(a),a}]$.
Note that we are giving the intuition behind our approach
and that many details are omitted.
For any $a \in A$, let $L = \{i_1, i_2, \dots, i_\ell\}$
be the set of impressions potentially matched with $a$ in \CaseOne. It follows that
$\Index(a)$ takes values in $L$ over the course of the algorithm.
Each of the impressions in $L \setminus \{i_1\}$ generates enough extra value
in the form of adaptive gain to increase the marginal gain of the previous impression.
The only expected marginal gain value that is lacking is the one associated with the $i_\ell$.
For this last impression, we consider a few different cases
and forward reference to the additional sources of marginal gain $Y_i$ and~$Z_i$
that arise in our lower bound for $\E[\ALG]$ in \Cref{lem:alg_lower_bound}.
The following argument is formalized by the \DistributeExcess mechanism in
\Cref{subsec:mechanism} for reallocating extra marginal gain.
Let $i'$ be the impression matched with $a$ in $\OPT$ (i.e., $a = a^*_{i'}$).
If $i'$ arrives before $i_\ell$, then $Y_{i'}$ is large and the mechanism
borrows from it to achieve enough value.
Now assume that $i'$ arrives after $i_\ell$.
If $i'$ is assigned in \CaseTwo, there is enough extra value to allocate to
$\E[\Gain_{i_\ell, a}]$ since this is the favorable case.
Otherwise, $i'$ is assigned in \CaseOne and $a$ is not one of the top candidates
in $\{a_1,a_2\}$.
If $a$ is not one of the top candidates because
$w_{i',a} < w_{\Index(a),a} - \delta M_{i'}$ (which makes the condition on line~7 false),
then the value of $Z_{i}$ is large enough to make up for this difference.
Otherwise, the sum $\AdaptiveGain_{i', a_1} + \AdaptiveGain_{i', a_2}$ is
large enough to cover the marginal gain of all three impressions
$i_\ell$, $\Index(a_1)$, and $\Index(a_2)$ since the \Cref{alg:stochastic_greedy}
accounts for this $2/3$ split in lines~14--15 when choosing $a_1$ and $a_2$.
Putting everything together, we show that all assignments in \CaseOne satisfy
$\E[\MarginalGain_i] \ge (1/2 + \varepsilon')\cdot(\E[\Gain_{i,a_1}] + \E[\Gain_{i,a_2}])$,
for some $\varepsilon' > 0$.

\section{Analysis of the Competitive Ratio}
\label{sec:analysis}

In this section we analyze the competitive ratio of \StochasticGreedy
and show that it breaks the $1/2$~barrier.
We start by presenting the high-level structure of our argument in~\Cref{subsec:proof_outline},
deferring the proofs of our two core lemmas (\Cref{lem:excess_and_xyz} and
\Cref{lem:lambda_gain}) to the following subsections.
In~\Cref{subsec:mechanism} we introduce a mechanism called~\DistributeExcess,
which we use in our analysis to systematically redistribute excess marginal gain.
We stress that \DistributeExcess exists solely for the sake of analysis
and is not a component of~\StochasticGreedy.
Then in~\Cref{subsec:main_proof} we prove \Cref{lem:lambda_gain}
by cases and show that \DistributeExcess
allocates enough excess marginal gain to every impression
for \StochasticGreedy to be 0.501-competitive.

\subsection{Outline of the Main Proof}
\label{subsec:proof_outline}

For every impression $i \in I$, we compare $\E[\MarginalGain_i]$ to the expected value
that $\StochasticGreedy$ could have achieved by assigning $i$ to $a_i^*$,
namely $\E[\Gain_{i,a_i^*}] = \E[(w_{i,a_i^*} - \MaxW_{a_i^*}^{t_i-1})^+]$.
To prove that greedy algorithms achieve a $1/2$ competitive ratio,
it usually suffices to show that $\E[\MarginalGain_i] \ge \E[\Gain_{i,a_i^*}]$.
Our algorithm \StochasticGreedy, however, is designed in such a way that this
condition is not necessarily satisfied for every impression.
Instead, we show that the \emph{sum of expected marginal gains} over all impressions
is significantly greater than this lower bound in aggregate.
Intuitively, impressions that beat this benchmark share their excess marginal
gain with other impressions so that at the end of the algorithm, every
impression contributes enough to exceed the standard $1/2$ competitive ratio.
One of the major technical contributions of this paper is our carefully
designed mechanism \DistributeExcess (\Cref{alg:distribute_excess}), which reallocates
the potential excess marginal gain of an assignment to guarantee a uniform lower
bound for every impression.

We begin by lower bounding the expected weight of the assignment that
\StochasticGreedy makes in such a way that
reveals three additional sources of potential excess gain that can be exploited.

\begin{lemma}
\label{lem:alg_lower_bound}
The expected weight of the assignment of \StochasticGreedy, namely $\E[\ALG]$, is at least
\[
  \frac{1}{2}\OPT + \frac{1}{2}\sum_{i \in I}
    \overbrace{\E\bracks*{\MarginalGain_i - \Gain_{i, a_i^*}}}^{X_i} +
    \overbrace{\E\bracks*{\MaxW_{a_i^*}^n - \MaxW_{a_i^*}^{t_i - 1}}}^{Y_i} +
    \overbrace{\E\bracks*{\parens*{\MaxW_{a_i^*}^{t_i-1} - w_{i,a_i^*}}^+}}^{Z_i}.
\]
\end{lemma}

\begin{proof}
We know that
$\E[\ALG] = \sum_{i \in I} \E[\MarginalGain_i] = \sum_{i \in I} X_i + \E[\Gain_{i,a_i^*}]$.
By the definition of the $(x)^+$ operator, we have
$\Gain_{i,a_i^*} = w_{i,a_i^*} - \MaxW_{a_i^*}^{t_i-1} 
  + \parens{\MaxW_{a_i^*}^{t_i-1} - w_{i,a_i^*}}^+$, which gives us the $Z_i$ term.
So far we have shown that
\[
  \E[\ALG] = \sum_{i \in I} X_i + \E\bracks*{w_{i,a_i^*} - \MaxW_{a_i^*}^{t_i-1}} + Z_i.
\]
On the other hand, we know 
$\E[\ALG] = \sum_{a \in A} \E[\MaxW_{a}^n]$ and
$\OPT = \sum_{i \in I} w_{i,a_i^*}$.
Writing
$w_{i,a_i^*} - \MaxW_{a_i^*}^{t_i-1}$ as
\[
  \parens*{w_{i,a_i^*} - \MaxW_{a_i^*}^n} + \parens*{\MaxW_{a_i^*}^n - \MaxW_{a_i^*}^{t_i-1}}
\]
yields with the $Y_i$ term.
Since $\sum_{i \in I} \E[\MaxW_{a_i^*}^n] \le \sum_{a \in A} \E[\MaxW_{a}^n] = \E[\ALG]$,
it follows that
\begin{align*}
  \E\bracks*{\ALG} &= 
    \sum_{i \in I} X_i + \E\bracks*{w_{i,a_i^*} - \MaxW_{a_i^*}^{n}} + Y_i + Z_i
    \ge \OPT - \E[\ALG] + \sum_{i \in I} X_i + Y_i + Z_i,
\end{align*}
which completes the proof.
\end{proof}

\noindent
The expectations $Y_i$ and $Z_i$ are nonnegative for all $i \in I$ since
the random variables $\MaxW_{a_i^*}^t$ are nondecreasing in $t$
and by the definition of the $(x)^+$ operator.
The expectations $X_i$, however, can sometimes be negative.

For the sake of analysis, we define an auxiliary variable $\Excess_i$ for each 
impression $i \in I$ and show how to assign its value at any given step of the algorithm.
We reiterate that the $\Excess_i$ variables are not actually used in \StochasticGreedy
and are only defined to help us prove the competitive ratio.
As noted above, the sum $X_i + Y_i + Z_i$ is not necessarily nonnegative.
Therefore, we introduce the mechanism~\DistributeExcess to systematically
redistribute excess marginal gain that is incurred over the course of the algorithm.
This allows us to assign values to all the $\Excess_i$ variables so that
$\sum_{i \in I} \Excess_i \le \sum_{i \in I} X_i + Y_i + Z_i$,
and more importantly,
for every impression $i \in I$, we have $\Excess_i \ge \lambda M_i$ for some
universal constant $\lambda > 0$.
In \Cref{thm:main_approximation}, we exploit these two properties of the $\Excess_i$
variables to prove that $\StochasticGreedy$ 
is at least $\frac{1 + \lambda}{2 + \lambda} > 1/2$ competitive.
Note that we call the routine of assigning values to the $\Excess_i$ variables a
mechanism (and not an algorithm) because it is only used in our analysis
as a means to argue about the aggregate excess marginal gain that
\StochasticGreedy produces.

For now, we abstract away the details of \DistributeExcess so that we can
understand its role in our analysis of the competitive ratio.
We present the mechanics of \DistributeExcess 
and the proof of the following lemma in \Cref{subsec:mechanism}.

\begin{lemma}
\label{lem:excess_and_xyz}
The mechanism $\DistributeExcess(\zeta,\gamma,\sigma)$ computes a value $\Excess_i$
for each impression $i \in I$ such that
$\sum_{i \in I} \Excess_i \le \sum_{i \in I} X_i + Y_i + Z_i$,
where the terms $X_i$, $Y_i$, and $Z_i$ are defined in
\Cref{lem:alg_lower_bound}.
Furthermore, for every $i \in I$,
at least a $\zeta$ fraction of $Z_i$ is distributed to $\Excess_i$.
\end{lemma}

The next lemma covers a variety of different cases
that \StochasticGreedy can encounter
and is the crux of our analysis.
In particular, \Cref{lem:lambda_gain} shows that we can use \DistributeExcess to uniformly lower bound each $\Excess_i$
variable in terms of the maximum expected gain $M_i$ when $i$ arrives.
We note that the inputs $\varepsilon, \delta$ to \StochasticGreedy
and $\zeta, \gamma, \sigma$ to \DistributeExcess
are intentionally left as variables so that we can optimize them retroactively
and so that the case analysis in the proof of \Cref{lem:lambda_gain} 
is simpler.
We discuss our approach and the supporting lemmas
for proving \Cref{lem:lambda_gain} in more detail in \Cref{subsec:main_proof}.

\begin{lemma}
\label{lem:lambda_gain}
For any $\varepsilon, \delta \ge 0$,
the mechanism $\DistributeExcess(\zeta,\gamma,\sigma)$
finds a value $\Excess_i \ge \lambda M_i$ for each impression $i \in I$,
where $\lambda$ is defined to be
$\lambda = \lambda(\varepsilon,\delta)
=\max_{0\le \zeta,\gamma,\sigma\le 1}
\min\{
  \frac{\varepsilon - 2\gamma - 2\sigma}{2},
  \frac{1-3\varepsilon-4\gamma-4\sigma}{4},
  \frac{2\zeta\delta - 3\varepsilon - 6\gamma - 6\sigma}{6},\allowbreak
  \frac{3-22\varepsilon}{19},
  \frac{6\zeta\delta - 1 - 18\varepsilon}{18},
  \frac{324(1-\varepsilon)^2 - 361\delta}{18468(1-\varepsilon)},
  \frac{324(1-\varepsilon)^2 - 361\delta}{18468(1-\varepsilon)} \times 18\sigma,
  \frac{2(1-\varepsilon)}{19},
  (1-\zeta)\frac{\delta}{1+\delta} \times \frac{6(1-\varepsilon)}{19},
  \frac{18(1-\varepsilon)}{19} \times \sigma,
  \frac{2\gamma}{1 + \delta} \times \frac{18(1-\varepsilon)}{19}
\}.$
In particular,
by setting
  $\varepsilon = 0.082$,
  $\delta = 0.445$,
  $\zeta = 0.955$,
  $\gamma=0.00337198$,
  and $\sigma=0.03362$,
we have $\lambda \ge 0.00400802$.
\end{lemma}

Now that we have presented our key prerequisite lemmas, we show how to
assemble them to prove our main result about \StochasticGreedy.

\begin{theorem}
\label{thm:main_approximation}
For any $\varepsilon, \delta \ge 0$,
the algorithm $\StochasticGreedy(\varepsilon,\delta)$ is
$\frac{1+\lambda}{2+\lambda}$-competitive, where
$\lambda = \lambda(\varepsilon, \delta)$ is defined in \Cref{lem:lambda_gain}.
In particular, if $\varepsilon = 0.082$ and $\delta = 0.445$,
then $\lambda \ge 0.00400802$ and \StochasticGreedy is $0.501$-competitive.
\end{theorem}

\begin{proof}
By combining \Cref{lem:alg_lower_bound}, \Cref{lem:excess_and_xyz},
and \Cref{lem:lambda_gain}, we know that
\begin{align*}
  \E\bracks*{\ALG} \ge \frac{1}{2}\OPT + \frac{1}{2}\sum_{i \in I} X_i + Y_i + Z_i
    \ge \frac{1}{2}\OPT + \frac{1}{2}\sum_{i \in I} \Excess_{i}
    \ge \frac{1}{2}\OPT + \frac{1}{2}\sum_{i \in I} \lambda M_i.
\end{align*}
For each impression we also have $M_i \ge \E[\Gain_{i,a_i^*}]$, which is at least
$\E[w_{i,a_i^*} - \MaxW_{a_i^*}^{t_i-1}] \ge \E[w_{i,a_i^*} - \MaxW_{a_i^*}^{n}]$.
Summing this lower bound over all impressions gives us
$\sum_{i \in I} M_i \ge \OPT - \E[\ALG]$.
Therefore, it follows that
$\E[\ALG] \ge \frac{1}{2}\OPT + \frac{\lambda}{2}(\OPT - \E[\ALG])$,
or equivalently $\E[\ALG] \ge \frac{1+\lambda}{2+\lambda} \OPT$.
\end{proof}

\subsection{Mechanism for Distributing Excess Marginal Gain}
\label{subsec:mechanism}

In this subsection we introduce the \DistributeExcess mechanism,
which reallocates the excess marginal gain
$\sum_{i \in I} X_i + Y_i + Z_i$ defined in \Cref{lem:alg_lower_bound}
to a set of auxiliary variables called $\Excess_i$.
This mechanism and the $\Excess_i$ variables are not part of the \StochasticGreedy algorithm,
and are used only to analyze the competitive ratio.
At any time during the online algorithm,
this mechanism assumes oracle access to the optimal assignment 
and assigns a value to each of the $\Excess_i$ variables in a way that
allows us to show that $\E[\ALG] \ge 0.501 \cdot \OPT$ for the current sequence of impressions.

At a high level, \DistributeExcess
mirrors the execution of \StochasticGreedy
and systematically redistributes the sum $\sum_{i \in I} X_i + Y_i + Z_i$
across all of the $\Excess_i$ variables.
Its execution path relies solely on deterministic quantities
in \StochasticGreedy (see \Cref{lem:deterministic}) and is thus independent of
its randomness.
The mechanism takes as input three parameters $\zeta, \gamma, \sigma$
representing allocation proportions that we optimize later.
The allocations of the $X_i$ and $Z_i$ variables are direct,
but the $Y_i$ variables are partitioned and distributed over time.
Before presenting \DistributeExcess,
we define a refinement of $Y_i$ to captures how it evolves.



\begin{definition}
\label{def:delta_vars}
For every $i \in I$,
we define the \emph{time sequence} of $Y_i$
(introduced in \Cref{lem:alg_lower_bound}) to be
\[
  Y_{i,t} = \begin{cases}
    0 & \text{if $0 \le t < t_i$,} \\
    \E\bracks*{\MaxW_{a_i^*}^t - \MaxW_{a_{i}^*}^{t_i - 1}} & \text{if $t_i \le t \le n$,}
\end{cases}
\]
We let $\Delta^t$ denote 
the \emph{backwards difference} for time sequence
values, implicitly defined as
$\Delta^t(Y_i) = Y_{i,t} - Y_{i,t-1}$.
\end{definition}


\SetAlgorithmName{Mechanism}{mechanism}{List of Mechanisms}
\begin{algorithm}[H]
\caption{Mechanism to populate the $\Excess_i$ variables.}
\label[mechanism]{alg:distribute_excess}
\DontPrintSemicolon

\SetKwProg{FunctionDistributeExcess}{\textbf{function} $\DistributeExcess(\zeta,\gamma,\sigma)$}{}{}

\FunctionDistributeExcess{}{
    Initialize $\Excess_i \gets 0$ for all $i \in I \cup \{0\}$
    \tcp*[r]{Recall that $0$ is the initial value of $\Index(a)$}

    \For{$t=1,2,\dots,|I|$}{
      Let $i$ be the impression that arrives at time $t$, i.e., $t_i = t$\;
      Increase $\Excess_i$ by $\zeta Z_{i}$
      \tcp*[r]{Distribute $\zeta Z_i$}

      \For{$a \in \{a_1,a_2\}$ \textnormal{(if $a_2$ does not exist, only consider $a_1$)}}{
        Let $i'$ be the impression for which $a = a_{i'}^*$
        \tcp*[r]{Distribute $\Delta^t(Y_{i'})$}

        \If{\textnormal{$i'$ arrived at or before time $t$}}{
          $K \gets \{i, i', \Index(a)\}$\;
          \For{$j \in K$}{
            Increase $\Excess_j$ by $\Delta^t(Y_{i'}) / |K|$
          }
        }
      }
      \If{$|B| \ge 2$}{
        \For(\tcp*[f]{Use the value of $\Index(a)$ at the beginning of time $t$}){$a \in \{a_1,a_2,a_i^*\}$}{
          Increase $\Excess_{\Index(a)}$ by $2/3 \cdot \AdaptiveGain_{i,a}$\;
        }
        Set $\Excess_i \gets \Excess_i + X_i - \sum_{a \in \{a_1,a_2,a_i^*\}} 2/3\cdot \AdaptiveGain_{i,a}$ \;
        \For(\tcp*[f]{Use the value of $\Index(a)$ at the end of time $t$}){$a \in \{a_1,a_2\}$}{
          \If{\textnormal{$\Index(a)$ remains unchanged after time $t$}}{
            Let $i'$ be the impression for which $a = a_{i'}^*$
            \tcp*[r]{Distribute $(1-\zeta) Z_{i'}$}
            Increase $\Excess_{i}$ by $(1-\zeta) Z_{i'}$\;
          }
        }
      }
      \Else{
        \lIf{\textnormal{$a_2$ exists}}{
          Increase $\Excess_{\Index(a_1)}$ and $\Excess_{\Index(a_2)}$ each by $\sigma M_i /2$
        }
        \lElse{
          Increase $\Excess_{\Index(a_1)}$ by $\sigma M_i$
        }
        Increase $\Excess_{\Index(a_i^*)}$ by $\gamma M_i$\;
        Set $\Excess_i \gets \Excess_i + X_i - (\sigma + \gamma)M_i$\;
      }
    }
}
\end{algorithm}

We conclude this subsection by proving~\Cref{lem:excess_and_xyz},
which guarantees that the mechanism performs
a valid reallocation.
In particular, we show that
the final assignments satisfy $\sum_{i \in I} \Excess_i \le \sum_{i \in I} X_i + Y_i + Z_i$.

\begin{proof}[Proof of \Cref{lem:excess_and_xyz}]
The second claim is immediate from line~5 of \Cref{alg:distribute_excess}, so
we focus on the first part of the statement.
At each time step $t = t_i$,
lines~13--15 and lines~21--24
change the sum $\sum_{j \in I \cup\{0\}} \Excess_j$ by exactly $X_i$.
Now we analyze lines 6--11 of \Cref{alg:distribute_excess}.
Assuming that $a_2$ exists, let $i_1'$ be the impression for which $a_1 = a_{i_1'}^*$
and 
let $i_2'$ be the impression for which $a_2 = a_{i_2'}^*$.
Since $a_1 \ne a_2$ we know $i'_1 \ne i'_2$.
Therefore, the change in $\sum_{j \in I \cup\{0\}} \Excess_j$
at time $t = t_i$ is at most $\Delta^t(Y_{i_1'}) + \Delta^t(Y_{i_2'}) = \sum_{j \in I} \Delta^t(Y_j)$.
If $a_2$ does not exist, we can also upper bound the change
in $\sum_{j \in I \cup\{0\}} \Excess_j$ at time $t=t_i$ by
$\Delta^t(Y_{i_1'}) = \sum_{j \in I} \Delta^t(Y_j)$.
Recall that we have $\sum_{t=1}^n \Delta^t(Y_i) = Y_i$ by \Cref{def:delta_vars}.
Summing over all time steps~$t$,
the total contribution of lines 6--11 at the end of \Cref{alg:distribute_excess}
to~$\sum_{i \in I \cup\{0\}} \Excess_i$ is at most $\sum_{i \in I} Y_i$.
Now we analyze the contribution of the~$Z_i$ variables.
The sum $\sum_{j \in I \cup \{0\}} \Excess_j$ is clearly increased at time $t=t_i$ by 
$\zeta Z_i$ on line~5.
Next, observe that the total contribution from lines~16--19
to the sum $\sum_{j \in I \cup\{0\}} \Excess_j$
over the course of the mechanism is at most
$(1-\zeta) \sum_{i \in I} Z_i$.
This follows from the facts that the conditional statement on line~17 evaluates to
true at most once for each $a \in A$,
and that each impression is matched to at most one advertiser in the optimal assignment.
Since all of these contributions are disjoint,
we have
$\sum_{j \in I \cup\{0\}} \Excess_j \le \sum_{i \in I} X_i + Y_i + Z_i$
at the end of \Cref{alg:distribute_excess}.
The claim follows because $\Excess_0$ is always increased by nonnegative amounts.
\end{proof}

\subsection{Lower Bounding the Excess Allocated to Each Impression}
\label{subsec:main_proof}

The rest of our analysis is devoted to proving~\Cref{lem:lambda_gain}.
We will need the following three inequalities in addition to~\Cref{lem:excess_and_xyz},
so we state them together for ease of reference and defer the longer proofs
to~\Cref{app:analysis}.
First, we show in \Cref{lem:marginal_gain_lower_bound} 
how the adaptive decisions in lines~23--24 of \Cref{alg:stochastic_greedy}
improve the expected marginal gain when assigning impressions.
We note that although there are other potential adaptive opportunities to
exploit in \Cref{alg:stochastic_greedy} (e.g., lines~41--42), 
we use adaptiveness in a very controlled way and limit its use to the most
beneficial parts of the algorithm.

\begin{restatable}[]{lemma}{MarginalGainLowerBound}
\label{lem:marginal_gain_lower_bound}
If impression $i$ is assigned in \CaseOne of \StochasticGreedy,
i.e., case $|B| \ge 2$, we have
\[
  \E\bracks*{\MarginalGain_i}
  \ge \frac{\E\bracks*{\Gain_{i,a_1}} + \E\bracks*{\Gain_{i,a_2}}}{2} + 
  \AdaptiveGain_{i,a_1} + \AdaptiveGain_{i,a_2}.
\]
\end{restatable}

The following two inequalities are derivatives of \Cref{lem:marginal_gain_lower_bound}
that show how $\AdaptiveGain_{i,a}$, $\E[\Gain_{i,a}]$,
and~$M_i$ relate to one another.
We note that the proof of \Cref{lem:adaptive_gain_upper_bound}
builds directly on the proof of~\Cref{lem:marginal_gain_lower_bound}.

\begin{restatable}[]{lemma}{AdaptiveGainUpperBound}
\label{lem:adaptive_gain_upper_bound}
For any advertiser $a \in A$ and impression $i \in I$,
  we have $\AdaptiveGain_{i,a} \le \frac{1}{12} \E[\Gain_{i,a}]$.
\end{restatable}

\begin{restatable}[]{lemma}{ExpectedGainLowerBounds}
\label{lem:expected_gain_lower_bounds}
For any advertiser $a \in B$ at time $t=t_i$ for impression $i$, we have
$\E[\Gain_{i,a}] \ge \frac{18(1-\varepsilon)}{19} M_i$.
\end{restatable}

\begin{proof}
This is a direct consequence of
\Cref{lem:adaptive_gain_upper_bound}
and the definition of the set $B$ in
line~11 of \Cref{alg:stochastic_greedy}.
\end{proof}

Now we present the proof of \Cref{lem:lambda_gain}, which
completes the analysis of \Cref{alg:distribute_excess}
and consequently the competitive ratio of \Cref{alg:stochastic_greedy}.
To show that $\Excess_i \ge \lambda M_i$ for every impression $i \in I$,
we first consider three different
scenarios that can occur when $i$ arrives.
For each of the three top-level cases,
we analyze a series of subcases,
all of which result in lower bounds for $\Excess_i$ that are a multiple of~$M_i$.
The branching structure of these cases is initially difficult to
discern, but it is somewhat unavoidable given the adaptivity of
\Cref{alg:stochastic_greedy} and the design of \Cref{alg:distribute_excess}.
We note, however, that the
subcases themselves are relatively easy to verify.
We present a distilled version of the subcases and their implications in~\Cref{fig:tree}.

\begin{figure}[H]
\begin{tcolorbox}[
        colback=white
    ]
\begin{longenum}
\itemsep0em

\vspace{-0.05cm}
\item Case 1: $i$ is assigned in lines 31--42
  \vspace{-0.15cm}
  \begin{longenum}
  \itemsep0em

    \item $a_i^* \not\in B' \cup C  \implies \frac{\varepsilon - 2\gamma - 2\sigma}{2}M_i$
    \item $a_i^* \in B'  \implies \frac{1-3\varepsilon - 4\gamma - 4\sigma}{4}M_i$
    \item $a_i^* \in C  \implies \frac{2\zeta \delta - 3\varepsilon - 6\gamma - 6\sigma}{6} M_i$
  \end{longenum}

  \vspace{-0.15cm}
  \item Case 2: $i$ is assigned in lines 13--30 and $a_i^* \in \{a_1,a_2\}$
        $\implies \frac{3-22\varepsilon}{19} M_i$

  \item Case 3: $i$ is assigned in lines 13--30 and $a_i^* \not\in \{a_1,a_2\}$

  \vspace{-0.20cm}
  \begin{longenum}
  \itemsep0em
    \item $w_{i,a_i^*} < w_{\Index(a_i^*), a_i^*} - \delta M_i$
         $\implies \frac{6\zeta\delta - 1 - 18\varepsilon}{18} M_i$
    \item $\E[\Gain_{i,a_i^*}] + 2/3 \cdot \AdaptiveGain_{i,a_i^*} 
              \le \E[\Gain_{i,a_j}] + 2/3\cdot\AdaptiveGain_{i,a_j}$ for $j \in \{1,2\}$
    \begin{longenum}
      \itemsep0em
      \item $\Index(a_1)$ or $\Index(a_2)$ changes after $t$
        $\implies \min\left\{\frac{324(1-\varepsilon)^2 - 361\delta}{18468(1-\varepsilon)}, 
              \frac{324(1-\varepsilon)^2 - 361\delta}{18468(1-\varepsilon)} \times 18\sigma
          \right\}M_i$
      \item $\Index(a_1)$ and $\Index(a_2)$ remain unchanged after time $t$
      \begin{longenum}
        \item Impression $i_1$ arrived at or before time $t$ where $a_1 = a_{i_1}^*$
            $\implies \frac{2(1-\varepsilon)}{19}M_i$
       \item Impression $i_2$ arrived at or before time $t$ where $a_2 = a_{i_2}^*$
            $\implies \frac{2(1-\varepsilon)}{19}M_i$
        \item Impressions $i_1$ and $i_2$ arrive after time $t$
        \begin{longenum}
          \item $w_{i_1, a_1}\hspace{-0.08cm} <\hspace{-0.06cm} w_{i, a_1} - \delta M_{i_1}$\hspace{-0.03cm} or \hspace{-0.03cm} $w_{i_2, a_2}\hspace{-0.08cm} <\hspace{-0.06cm} w_{i, a_2} - \delta M_{i_2}$
            $\hspace{-0.09cm} \implies\hspace{-0.08cm}  \min\left\{
              (1\hspace{-0.06cm}-\hspace{-0.06cm}\zeta)\frac{\delta}{1+\delta}\hspace{-0.06cm} \times\hspace{-0.06cm} \frac{6(1-\varepsilon)}{19},
                  \frac{18(1-\varepsilon)}{19}\hspace{-0.06cm} \times\hspace{-0.06cm} \sigma
              \right\}\hspace{-0.03cm} M_i$
          \item $w_{i_1, a_1} \ge w_{i, a_1} - \delta M_{i_1}$ and
                $w_{i_2, a_2} \ge w_{i, a_2} - \delta M_{i_2}$
                $\implies \min\left\{
                  \frac{2\gamma}{1 + \delta} \hspace{-0.05cm} \times\hspace{-0.05cm} 
                        \frac{18(1-\varepsilon)}{19},
                  \frac{18(1-\varepsilon)}{19} \hspace{-0.05cm}\times\hspace{-0.05cm} \sigma
                    \right\} M_i$
        \end{longenum}
      \end{longenum}
    \end{longenum}
  \end{longenum}
\end{longenum}
\end{tcolorbox}
  \caption{Branching structure of the subcases for lower bounding $\Excess_i$ in the proof of \Cref{lem:lambda_gain}.}
  \label{fig:tree}
\end{figure}

\begin{proof}[Proof of \Cref{lem:lambda_gain}]
It suffices to show that for any $0 \le \zeta,\gamma,\sigma \le 1$,
mechanism $\DistributeExcess(\zeta,\gamma,\sigma)$
increases $\Excess_i$ by at least
$\min\set{
  \frac{\varepsilon - 2\gamma - 2\sigma}{2},
  \frac{1-3\varepsilon - 4\gamma - 4\sigma}{4},
  \frac{2\zeta \delta - 3\varepsilon - 6\gamma - 6\sigma}{6},
  \frac{3-22\varepsilon}{19},
  \frac{6\zeta \delta - 1 - 18\varepsilon}{18},
  \frac{324(1-\varepsilon)^2 - 361\delta}{18468(1-\varepsilon)},
  \frac{324(1-\varepsilon)^2 - 361\delta}{18468(1-\varepsilon)} \times 18\sigma,
  \frac{2(1-\varepsilon)}{19},
  (1-\zeta)\frac{\delta}{1+\delta} \times \frac{6(1-\varepsilon)}{19},
  \frac{18(1-\varepsilon)}{19} \times \sigma,
  \frac{2\gamma}{1+\delta} \times \frac{18(1-\varepsilon)}{19}
} M_i$.
We start by observing that $\Excess_i$ might change during
\Cref{alg:distribute_excess} for a variety of reasons.
Since the sequence $Y_{i,t}$ is nonnegative and nondecreasing in $t$,
the updates to the $\Excess_j$ variables in line~11 are nonnegative.
Similarly, $Z_i$ is nonnegative so the changes in line~5 and line~19 are
nonnegative.
The $\Excess$ variables also do not decrease in line~14
or lines 21--23 because
the $\AdaptiveGain$ and $M_i$ variables are nonnegative.
The only place where $\Excess_i$ might be reduced is in line~15 or line~24 at $t=t_i$.
In particular, this happens for small or negative values of~$X_i$.
In this proof, instead of tracking all changes to $\Excess_i$, we bound
this one-time reduction to $\Excess_i$ and show that $\Excess_i$
is increased enough elsewhere to compensate for this potential decrease.
We consider three main cases and prove each separately:
(1) impression $i$ is assigned in lines \CaseTwo of \Cref{alg:stochastic_greedy},
(2) $i$ is assigned in \CaseOne and $a_i^* \in \{a_1,a_2\}$,
and (3) impression $i$ is assigned in \CaseOne and $a_i^* \not\in \{a_1,a_2\}$.

\paragraph{Case 1 ($i$ is assigned in \CaseTwo).}
We start by proving the claim in the simplest case when $i$
is assigned in \CaseTwo of \Cref{alg:stochastic_greedy}.
Since $\AdaptiveGain$ is always nonnegative, $B'$ is a subset of $B$.
According to the else condition on line~31 of \Cref{alg:stochastic_greedy},
the set $B$, and thus $B'$, contains at most one advertiser.
This is why we say the only advertiser in~$B'$ on line 38.
Recall that $X_i = \E[\MarginalGain_i] - \E[\Gain_{i,a_i^*}]$.
If $a_i^* \not\in B' \cup C$, then $\E[\Gain_{i,a_i^*}] < (1-\varepsilon)M_i$
by the definition of sets $B'$ and $C$.
On the other hand, 
$\argmax_{a \in A} \E[\Gain_{i,a}]$ will be selected as
one of the advertisers to which $i$ is assigned,
and it achieves a gain of $M_i$ by definition.
If there is a second choice (i.e., $a_2$ exists), its gain is at least
$(1-\varepsilon)M_i$ by the definition of the sets $B'$ and $C$.
Therefore, $\E[\MarginalGain_i]$ is at least
$(1-\sfrac{\varepsilon}{2})M_i$, which implies that
$X_i \ge \sfrac{\varepsilon}{2} M_i$.
It follows from line~24 of \Cref{alg:distribute_excess} that
$\Excess_i$ is increased by at least
$\frac{\varepsilon - 2\gamma - 2\sigma}{2} M_i$,
which proves the claim.

Next, we consider the scenario $a_i^* \in B'$.
In this subcase,
\Cref{alg:stochastic_greedy} selects $a_i^*$ as one of at most
two candidates for $i$.
This potential assignment increases $Y_i$ at time $t$ by at least
$\E[\Gain_{i,a_i^*}]/2$, which implies that
$\Delta^t(Y_i) \ge \E[\Gain_{i,a_i^*}]/2 \ge (1-\varepsilon) M_i / 2$.
Observe that this situation causes \Cref{alg:distribute_excess}
to increase $\Excess_i$ in line~11. Since $i' = i$ in this subcase,
we have $|K| \le 2$, which means
$\Excess_i$ is increased by at least $(1-\varepsilon)M_i / 4$.
On the other hand, similar to the argument above, we can show that
the change in $\Excess_i$ in line~24 of \Cref{alg:distribute_excess}
is at least
$(1-\sfrac{\varepsilon}{2})M_i - M_i - (\gamma + \sigma)M_i
= -\frac{\varepsilon + 2\gamma + 2\sigma}{2} M_i$.
Therefore, at the end of \Cref{alg:distribute_excess}, the value of
$\Excess_i$ is at least
$\frac{1-\varepsilon}{4}M_i - \frac{\varepsilon + 2\gamma + 2\sigma}{2}M_i
=\frac{1-3\varepsilon - 4\gamma - 4\sigma}{4}M_i$,
which again proves the claim.

To conclude Case~1, we assume that $a_i^* \in C$.
Similar to the argument in the previous paragraph, we can show that
the change in $\Excess_i$ is at least
$-\frac{\varepsilon + 2\gamma + 2\sigma}{2}M_i$ in line~24 of \Cref{alg:distribute_excess}.
Since $a_i^* \in C$, we know that
$w_{i,a_i^*} < w_{\Index(a_i^*),a_i^*} - \delta M_i$.
Observe that $\Index(a_i^*) \ne 0$, for otherwise we would have $w_{i,a_i^*} < 0$.
Therefore, impression $\Index(a_i^*)$ arrived before $i$ and was assigned to $a_i^*$
with probability at least $1/3$.
This potential assignment of
$\Index(a_i^*)$ to $a_i^*$ at time $t_{\Index(a_i^*)}$
implies that
$Z_i \ge (w_{\Index(a_i^*),a_{i}^*} - w_{i,a_i^*})/3 > \delta M_i / 3$
by the definitions of $Z_{i}$ and $C$.
\Cref{lem:excess_and_xyz} implies that a $\zeta$ fraction of $Z_i$ is
distributed to $\Excess_i$, so
when \Cref{alg:distribute_excess} ends, 
$\Excess_i$ is at least
$\zeta Z_i - \frac{\varepsilon + 2\gamma + 2\sigma}{2}M_i
\ge \frac{\zeta \delta}{3} M_i - \frac{\varepsilon + 2\gamma + 2\sigma}{2}M_i
=\frac{2\zeta \delta - 3\varepsilon - 6\gamma - 6\sigma}{6} M_i$,
as desired.

\paragraph{Case 2 ($i$ is assigned in \CaseOne and $a_i^* \in \{a_1,a_2\}$).}
Now we consider the case where $i$ is assigned in \CaseOne of \Cref{alg:stochastic_greedy}
and $a_i^*$ is equal to either $a_1$ or $a_2$.
We show that $\Excess_i$ does not decrease too much in line~15 of
\Cref{alg:distribute_excess},
and then we lower bound its increments on other occasions.
Using \Cref{lem:marginal_gain_lower_bound}, we know
$\E[\MarginalGain_i]$ is at least
$(\E[\Gain_{i,a_1}] + \E[\Gain_{i,a_2}])/2 + \AdaptiveGain_{i,a_1} + \AdaptiveGain_{i,a_2}$.
Since $a_i^* \in \{a_1,a_2\}$, the change in $\Excess_i$ in line~15
at time $t=t_i$ is at least
\begin{align}
\label{eqn:x_drop}
  &\hspace{-0.10cm} X_i - \sum_{a \in \{a_1,a_2\}} 2/3 \cdot \AdaptiveGain_{i,a} =
  \E\bracks*{\MarginalGain_i} - \E\bracks*{\Gain_{i,a_i^*}}
    - \sum_{a \in \{a_1,a_2\}} 2/3 \cdot \AdaptiveGain_{i,a} \notag\\
  &\hspace{0.25cm}\ge
  \parens*{\sum_{a \in \{a_1,a_2\}}
    1/2 \cdot \E\bracks*{\Gain_{i,a}}
    + \AdaptiveGain_{i,a}
  }
  - \E\bracks*{\Gain_{i,a_i^*}}
  - \sum_{a \in \{a_1,a_2\}} 2/3 \cdot \AdaptiveGain_{i,a} \notag\\
  &\hspace{0.25cm}=
  \frac{1}{2}\left(\E\bracks*{\Gain_{i,a_1}} + 2/3 \cdot \AdaptiveGain_{i,a_1}
    +   \E\bracks*{\Gain_{i,a_2}} + 2/3 \cdot \AdaptiveGain_{i,a_2}\right)
      - \E\bracks*{\Gain_{i,a_i^*}}\\
  &\hspace{0.25cm}\ge (1-\varepsilon) M_i - M_i = -\varepsilon M_i, \notag
\end{align}
where the last inequality holds because $a_1$ and $a_2$ are both in the set $B$
and because $\E[\Gain_{i,a_i^*}] \le M_i$.

Now, since $a_i^*$ is the same as $a_1$ or $a_2$, variable $Y_i$ increases
by at least $\E[\Gain_{i,a_i^*}]/3$ at time $t$ because $1/3$ is a lower bound
on the probability of assigning $i$ to $a_i^*$.
Furthermore, \Cref{lem:expected_gain_lower_bounds}
implies $\E[\Gain_{i,a_i^*}] \ge \frac{18(1-\varepsilon)}{19} M_i$
because $a_i^* \in B$.
This means that at time $t=t_i$ on line~11 of \Cref{alg:distribute_excess},
$\Excess_i$ is increased by at least
$\Delta^t (Y_i) / |K| \ge \E[\Gain_{i,a_i^*}]/6 \ge \frac{3(1-\varepsilon)}{19} M_i$
since we have $i=i'$ and $|K| \le 2$.
Therefore, we conclude that at the end of \Cref{alg:distribute_excess},
the value $\Excess_i$ is at least
$\frac{3(1-\varepsilon)}{19} M_i - \varepsilon M_i = \frac{3-22\varepsilon}{19}M_i$,
which proves the claim for Case~2.

\paragraph{Case 3 ($i$ is assigned in \CaseOne and $a_i^* \not\in \{a_1,a_2\}$).}
The final case is when~$i$ is assigned in \CaseOne
and $a_i^* \not\in \{a_1,a_2\}$. We first bound the reduction of
$\Excess_i$ in line~15 of \Cref{alg:distribute_excess} at time $t=t_i$, and then we
prove it is increased enough in other occasions.
Like in Case~2, we start by applying~\Cref{lem:marginal_gain_lower_bound}.
The new idea we can use here is that since $a_1$ and $a_2$ have been selected
as the top two choices in the set~$B$ (lines~14--15 of \Cref{alg:stochastic_greedy})
and $a_i^*$ has not been chosen, at least one of the following inequalities
holds:
$w_{i,a_i^*} < w_{\Index(a_i^*),a_i^*} - \delta M_i$
or
$\E[\Gain_{i,a_i^*}] + 2/3\cdot \AdaptiveGain_{i,a_i^*} \le
 \E[\Gain_{i,a_j}] + 2/3\cdot \AdaptiveGain_{i,a_j}$, for both $j \in \{1,2\}$.
We start by proving the claim in the first scenario.
To do this, we need to introduce the new
$2/3\cdot\AdaptiveGain_{i,a_i^*}$ term into \Cref{eqn:x_drop}
from Case~2 to address the fact that $a_i^* \not\in \{a_1,a_2\}$.
Working from~\Cref{eqn:x_drop},
we can say that the change in $\Excess_i$ in line~15 of \Cref{alg:distribute_excess}
is at least
\[
  -\varepsilon M_i - \frac{2}{3} \cdot \AdaptiveGain_{i,a_i^*}
  \ge
  -\varepsilon M_i - \frac{2}{3} \cdot \frac{\E\bracks*{\Gain_{i,a_i^*}}}{12}
  \ge
  -\parens*{\frac{1}{18} + \varepsilon} M_i,
\]
where the first inequality holds by \Cref{lem:adaptive_gain_upper_bound}.
Since we first assume
$w_{i,a_i^*} < w_{\Index(a_i^*),a_i^*} - \delta M_i$,
impression $\Index(a_i^*)$ exists (i.e., it is not zero).
We know that impression $\Index(a_i^*)$ arrived before $i$ and is assigned
to~$a_i^*$ with probability at least $1/3$ in \Cref{alg:stochastic_greedy}.
Therefore, it follows that
$Z_i \ge (w_{\Index(a_i^*),a_i^*} - w_{i,a_i^*})/3 > \delta M_i/3$.
Applying~\Cref{lem:excess_and_xyz},
we know \Cref{alg:distribute_excess} increases
$\Excess_i$ on line~5 by at least $ \zeta \delta M_i / 3$ upon termination.
Therefore, the final value of $\Excess_i$ is at least
$(\frac{\zeta \delta}{3} - \frac{1}{18} - \varepsilon)M_i =
\frac{6\zeta\delta - 1 - 18\varepsilon}{18} M_i$,
which proves the claim for the first scenario.

To complete the analysis for Case~3,
we focus on the second subcase where we assume for $j \in \{1,2\}$ that
\begin{align*}
\label{eqn:case_3_2}
  \E\bracks*{\Gain_{i,a_j}} + \frac{2}{3} \cdot \AdaptiveGain_{i,a_j}
  \ge
  \E\bracks*{\Gain_{i,a_i^*}} + \frac{2}{3} \cdot \AdaptiveGain_{i,a_i^*}.
\end{align*}
To address the assumption that $a_i^* \not\in \{a_1,a_2\}$,
we adapt \Cref{eqn:x_drop} as follows 
and then combine it with the inequalities above
to get the following lower bound on the change of $\Excess_i$ in line~15 of
\Cref{alg:distribute_excess}:
\[
  \frac{1}{2} \left(\sum_{j \in \{1,2\}} \E\bracks*{\Gain_{i,a_j}} + \frac{2}{3}\cdot \AdaptiveGain_{i,a_j} \right) - \E\bracks*{\Gain_{i,a_i^*}} - \frac{2}{3} \cdot \AdaptiveGain_{i,a_i^*}
  \ge 0.
\]
Therefore, $\Excess_i$ is not reduced in line~15 of \Cref{alg:distribute_excess}.
To complete Case~3, it suffices to show that 
$\Excess_i$ is increased enough in other places.

We note that in line~17 of \Cref{alg:stochastic_greedy},
$\Index(a_1)$ and $\Index(a_2)$ are set to $i$
and $\Color_{a_1}$ and $\Color_{a_2}$ are set to $\blue$.
For now, we assume that at least one of these two indices changes after time $t$.
(We prove the claim later if this assumption does not hold.)
Let $t'$ be the first time that this happens and let $i'$ be the impression that
arrives at time $t'$.
Without loss of generality, we assume that $\Index(a_1)$ is the one among
these two that changes from $i$ to $i'$ at time $t'$.
(The following argument also holds even if both of them change at $t'$.)
At time $t'$, line~14 of \Cref{alg:distribute_excess} increases $\Excess_i$ by
$2/3\cdot\AdaptiveGain_{i',a_1}$
because $\Index(a_1)$ is equal to $i$ before it is set to $i'$.
In the time period $[t+1,t'-1]$, the indices of $a_1$ and $a_2$ remain
unchanged, and thus they are always active.
Note that if $t' = t+1$, the time period is empty and the claim about their
active state still holds.
Since $a_1$ is one of the two advertisers that \Cref{alg:stochastic_greedy}
selects for $i'$, 
we know~$a_1$ is in the set~$B$ at time $t'$,
which further implies $w_{i',a_1} \ge w_{i,a_1} - \delta M_{i'}$.
Therefore, the if condition on line 7 of \Cref{alg:stochastic_greedy} is true
for $a_1$ at time $t'$, and
$\AdaptiveGain_{i',a_1}$ is set to
$(\E[\Gain_{i,a}]/3 - (w_{i,a_1} - w_{i',a_1})^+/3 - S_{a_1})^+/12$.
Recall that $\Excess_i$ is increased at time $t'$ by
\begin{align}
\label{eqn:three_terms}
  \frac{2}{3} \cdot \AdaptiveGain_{i',a_1} = \parens*{
    \frac{\E\bracks*{\Gain_{i,a_1}}}{54}
    - \frac{\parens*{w_{i,a_1} - w_{i',a_1}}^+}{54}
    - \frac{S_{a_1}}{18}
  }^+.
\end{align}

We bound each of the three terms in \Cref{eqn:three_terms} separately.
Since $a_1$ is in set $B$ when both $i$ and $i'$ arrive, we have
$\E[\Gain_{i,a_1}] \ge \frac{18(1-\varepsilon)}{19} M_i$
and
$\E[\Gain_{i',a_1}] \ge \frac{18(1-\varepsilon)}{19} M_{i'}$
by \Cref{lem:expected_gain_lower_bounds}.
The term
$(w_{i,a_1} - w_{i',a_1})^+$ is zero if the weights satisfy $w_{i,a_1} \le w_{i',a_1}$.
If we have $w_{i,a_1} > w_{i',a_1}$, then
$\E[\Gain_{i,a_1}] \ge \E[\Gain_{i',a_1}]$ because $i$ arrives before
$i'$ and also has a larger weight to advertiser $a_1$.
Since $M_i \ge \E[\Gain_{i,a_1}]$, we also have $M_i \ge \frac{18(1-\varepsilon)}{19} M_{i'}$.
Applying the inequality
$w_{i,a_1} - w_{i',a_1} \le \delta M_{i'}$
from the definition of set $B$
shows that $\Excess_i$ is increased by at least
\[
  \parens*{\frac{18(1-\varepsilon)M_i}{54 \cdot 19}
  - \frac{19 \delta M_i}{54 \cdot 18(1-\varepsilon)}
  - \frac{S_{a_1}}{18}
  }^+
  =
  \parens*{
    \frac{324(1-\varepsilon)^2 - 361\delta}{18468(1-\varepsilon)}M_i
    -
    \frac{S_{a_1}}{18}
  }^+.
\]

Now we focus on lower bounding $S_{a_1}$.
At time $t'$, $S_{a_1}$ is the expected sum of $M_{i''}$
for every impression~$i''$
that has been assigned to $a_1$ in \CaseTwo of \Cref{alg:stochastic_greedy}
during the time period $[t+1,t'-1]$.
For each of these impressions,
\Cref{alg:distribute_excess} increases $\Excess_i$ in~lines 21-22 by
$\sigma M_{i''}/2$ or $\sigma M_{i''}$ 
depending on the existence of $a_2$ for $i''$.
Note that this is consistent with how \Cref{alg:stochastic_greedy} increases $S_{a_1}$.
Therefore, $\Excess_i$ is increased in total by at least
$\parens*{\frac{324(1-\varepsilon)^2 - 361\delta}{18468(1-\varepsilon)}M_i - \frac{S_{a_1}}{18}}^+ + \sigma S_{a_1}$
This lower bound proves the claim
because its minimum occurs when either
$S_{a_1}$ or the expression in the $(x)^+$ operator is zero.
In both cases, the lower bound is at least $\lambda M_i$.

We have reached the final step of Case~3 where we consider the scenario when
$\Index(a_1)$ and $\Index(a_2)$ both remain unchanged after time $t$,
and thus are active until the end of the algorithm.
If for some~$a \in \{a_1,a_2\}$, impression $i'$
(the impression with $a = a_{i'}^*$) arrived at or before time $t=t_i$,
we can lower bound the increase in $\Excess_i$ similar to in Case~2.
Since $i$ is assigned to $a$ with probability at least $1/3$,
$\Delta^t (Y_{i'})\ge \E[\Gain_{i,a}]/3$.
\Cref{alg:distribute_excess} increases $\Excess_i$ by at least
one third of this amount in line~11.
Since $a \in B$, we also know that
$\E[\Gain_{i,a}] \ge \frac{18(1-\varepsilon)M_i}{19}$ by \Cref{lem:expected_gain_lower_bounds}.
Therefore, $\Excess_i$ is increased by at least
$\frac{2(1-\varepsilon)}{19} M_i$, which proves the claim.

Now we consider the case where the impressions $i_1$ and $i_2$ arrive after time $t$,
where $i_1$ and $i_2$ are defined such that $a_1 = a_{i_1}^*$ and $a_2 = a_{i_2}^*$.
If $w_{i_1,a_1} < w_{i,a_1} - \delta M_{i_1}$, then we prove the claim as follows.
First, let $\Delta_{w}$ be the left-hand side of the inequality
$w_{i,a_1} - w_{i_1,a_1} > \delta M_{i_1}$.
We note that $\E[\Gain_{i_1,a_1}]$ is at least $\E[\Gain_{i,a_1}] - \Delta_w - S$, where $S$ is the value of $S_{a_1}$ at the
time $t'$ when $i_1$ arrives.
This lower bound holds because $\Delta_w$ compensates for how much smaller
$w_{i_1, a_1}$ is compared to $w_{i,a_1}$ and $S$ is
an upper bound on the total marginal gains of the edges assigned to $a_1$ between
the times between the arrivals of $i$ and $i_1$.
By definition, $M_{i_1} \ge \E[\Gain_{i_1,a_1}]$.
Using the assumption $\Delta_{w} > \delta M_{i_1}$, we have
$\Delta_w > \delta \E[\Gain_{i_1,a_1}] \ge \delta(\E[\Gain_{i,a_1}] - \Delta_w - S)$.
Applying \Cref{lem:expected_gain_lower_bounds}, we know that
$\E[\Gain_{i,a_1}] \ge \frac{18(1-\varepsilon)}{19} M_i$, which implies
$\Delta_w > \delta(\frac{18(1-\varepsilon)}{19}M_i - \Delta_w - S)$.
It follows that
$\Delta_w > \frac{\delta}{1+\delta}(\frac{18(1-\varepsilon)}{19}M_i - S)^+$.
Note that we can introduce the $(x)^+$ operator because $\Delta_w > \delta M_{i_1}$ is positive.
Since~$i$ arrived before $i_1$ and is assigned to $a_1$ with probability at least
$1/3$, variable $Z_{i_1}$ is at least $\Delta_w / 3$.
\Cref{alg:distribute_excess} increases $\Excess_i$ by exactly
$(1-\zeta) Z_{i_1}$ on line~19 because of our assumption that $\Index(a_1)$ remains unchanged.
Therefore, 
$\Excess_i$ is greater than
$(1-\zeta)\frac{\delta}{1+\delta}(\frac{6(1-\varepsilon)}{19} M_i - \frac{S}{3})^+ + \sigma S
\ge \lambda M_i$ at the end of the mechanism, which proves the claim.
The subcase when $w_{i_2,a_2} < w_{i,a_2} - \delta M_{i_2}$ follows similarly.

To complete the proof, we now assume
$w_{i_1,a_1} \ge w_{i,a_1} - \delta M_{i_1}$ and
$w_{i_2,a_2} \ge w_{i,a_2} - \delta M_{i_2}$.
If one of $i_1$ or~$i_2$ is assigned in \CaseOne,
the proof of the claim is identical to the part above starting near~\Cref{eqn:three_terms}
where we showed the claim for the scenario that at least one of
$\Index(a_1)$ or $\Index(a_2)$ changes after time $t$.
To see this, observe that if $i_1$ is assigned in \CaseOne, then
\Cref{alg:distribute_excess} increases $\Excess_{\Index(a_{i_1}^*)} = \Excess_i$ by
$2/3 \cdot \AdaptiveGain_{i_1, a_1}$ on line~14 at time $t_{i_1}$.
Otherwise, $\Excess_i$ is increased in line~23 by 
$\gamma(M_{i_1} + M_{i_2})$ when $i_1$ and $i_2$ arrive
since $\Index(a^*_1)$ and $\Index(a^*_2)$ are still equal to $i$.
Like in the previous paragraph, we can show that
$M_{i_1} \ge \E[\Gain_{i_1,a_1}] \ge \E[\Gain_{i,a_1}] - \Delta_{w}^1 - S'$,
where $\Delta_w^1 = (w_{i,a_1} - w_{i_1,a_1})^+$ and $S'$ is the value of $S_{a_1}$ when $i_1$ arrives.
By noting that $\Delta^1_w \le \delta M_{i_1}$
and applying \Cref{lem:expected_gain_lower_bounds},
we have the inequality $M_{i_1} \ge \frac{1}{1+\delta} (\frac{18(1-\varepsilon)}{19}M_i - S')^+$.
Observe that we can apply the $(x)^+$ operator because $M_{i_1}$ is nonnegative.
We can derive an analogous lower bound for $M_{i_2}$ by replacing $\Delta_{w}^1$ with $\Delta_{w}^2$
and $S'$ with $S''$.
Therefore, $\Excess_i$ is increased by at least
$\gamma \frac{1}{1+\delta}(\frac{18(1-\varepsilon)}{19}M_i - S')^+ + \sigma S'
+ \gamma \frac{1}{1+\delta}(\frac{18(1-\varepsilon)}{19}M_i - S'')^+ + \sigma S''$.
This lower bound proves the claim because its minimum occurs when either one of the
$(x)^+$ terms are zero, or both $S'$ and $S''$ are zero.
This completes Case~3 and therefore concludes the proof of \Cref{lem:lambda_gain}.
\end{proof}


\section{Conclusion}
\label{sec:conclusion}

We give the first algorithm for online weighted bipartite
matching with competitive ratio greater than~$1/2$
(under the free disposal assumption), resolving a central open problem in
the literature of online algorithms since the seminal work of Karp et
al.~\cite{karp1990optimal} thirty years ago.
Given the hardness result of Kapralov et al.~\cite{kapralov2013online},
our algorithm can be seen as strong evidence that solving the weighted bipartite
matching problem is strictly easier than submodular welfare maximization
in online settings.
Our main technical contributions in this work include a novel method for
making adaptive decisions that is amenable to analysis,
using the expectation of random variables over all possible branches of the randomized algorithm
to force key variables of the algorithm to be deterministic,
and a mechanism that we design solely for the sake of analysis
to systematically reallocate extra marginal gain that the algorithm produces.

\bibliographystyle{alpha}
\bibliography{references}

\newpage
\appendix

\section{Missing Analysis from \Cref{sec:algorithm}}
\label{app:algorithm}

\subsection{Proof of \Cref{lem:deterministic}}
\label{app:deterministic}

\Deterministic*

\begin{proof}
We proceed by induction on $t$. At the beginning of \Cref{alg:stochastic_greedy}
when $t=0$, all variables are initialized deterministically.
Assuming the claim as the induction hypothesis,
we proceed by analyzing the state of all the variables at time $t \ge 1$.
Recall that impression $i$ arrives at time $t=t_i$ and is predetermined by
the arrival order.
We begin by considering lines 4--12.
The maximum expected gain $M_i$ is an expected value over all branches of the
randomized algorithm up to time $t$ and is deterministic by definition.
For each $a \in A$, the current values of
$\Color(a)$, $\Index(a)$, and $S_a$ are deterministic by the induction hypothesis.
Therefore, all values of $\AdaptiveGain_{i,a}$ and the set
of candidates $B$ at time $t$ are also deterministic.

If we have $|B| \ge 2$, then the algorithm executes \CaseOne.
In this case, the advertisers $a_1$ and $a_2$ are the maximizers of
deterministic quantities and therefore deterministic themselves, assuming ties
are broken lexicographically.
The updates that occur in lines 16--21 do not rely on any randomness since the
value of $\Partner(a)$ in line~18 is fixed by the induction
hypothesis.
The branching in lines~22--30 possibly depend on the random values of $\rv{R}_i$
and $\Mark_{a_k}$, but in all of these conditional statements, only the
assignment of impression $i$ and updates to the variables $\Mark_{a_1}$ and
$\Mark_{a_2}$ are made.
In the second case, if $|B| \le 1$, then the algorithm executes \CaseTwo.
The only randomness here is the assignment of $i$ to either $a_1$ or $a_2$.
Therefore, the claim holds for all time steps $t$ by induction.
\end{proof}

\subsection{Proof of \Cref{lem:expected_gain_values}}

\ExpectedGainValues*

\begin{proof}
First,
recall that \Cref{alg:stochastic_greedy} is essentially deterministic except
for the impression assignments and the values of $\Mark_{a}$. To be specific,
\Cref{lem:deterministic} shows that the values of $\AdaptiveGain_{i,a}$
and the top candidates $a_1$ and~$a_2$ are deterministic and
  depend solely on the underlying instance
and the arrival order of the impressions.
Furthermore, recall that $\Gain_{i,a} = (w_{i,a} - \MaxW_{a}^{t_i - 1})^+$.
Therefore, it suffices to maintain the probability mass function for the
random variables $\MaxW_{a}^t$ at each time step of the algorithm.

We proceed by induction on $t$.
Let $W_{a}^t = \{w_{0,a}, w_{i_1,a}, w_{i_2,a}, \dots, w_{i_{t-1},a}\}$ denote the set
of possible weights assigned to advertiser $a$ at the beginning of time step $t \ge 1$,
and recall that $w_{0,a} = 0$.
Note that we use~$i_t$ to denote the impression that arrives at time $t$.
Next, let
$D_{a}^t : W_{a}^t \rightarrow \R_{\ge 0}$ be the probability mass function for
the random variable $\MaxW_{a}^t$ at the beginning of iteration $t$.
Since the state of advertiser $a$, namely $\Mark_a$, is randomized,
we refine the distribution $D_{a}^t$ into three conditional distributions. 
Let $p_{a,2}^t$ be the probability that $\Mark_a = 2$ at the beginning of
time $t$, and let $D_{a,2}^t$ be the distribution for $\MaxW_{a}^t$ given
that $\Mark_a = 2$ at the beginning of time $t$. The probabilities
$p_{a,1}^t, p_{a,0}^t$ and conditional distributions $D_{a,1}^t, D_{a,0}^t$
are defined similarly.
We note that if a distribution is not well-defined
because $p_{a,j}^t = 0$ for some $j \in \{0,1,2\}$,
our analysis still holds since because these update rules simulate all branches
of the randomized algorithm and always pull information from reachable states.
Note also that we have
$D_{a,t} = \sum_{j=0}^2 p_{a,j}^t D_{a,j}^t$
by the law of total probability if we treat the addition and
scalar multiplication of probability distributions like vectors.

No impressions have been assigned at the beginning of time $t=1$,
so for every advertiser $a \in A$ the probabilities are set to
$p_{a,0}^1 = 1, p_{a,1}^1 =0, p_{a,2}^1 = 0$
and the distributions satisfy $D_{a,j}^1(0) = 1$ for all $j \in \{0,1,2\}$.
This completes the base case when $t=1$,
so now assume $t \ge 1$ and that we have computed $p_{a,j}^{t}$
and $D_{a,j}^{t}$ for all $a \in A$ and $j \in \{0,1,2\}$.
We show how to compute all of these quantities at time $t+1$.
Let $a_1$ and $a_2$ be the top candidates at time $t$,
and let $i = i_t$.
(If $a_2$ does not exist, we can ignore this term.)
For every advertiser $a \in A \setminus\{a_1,a_2\}$, the state of $\Mark_{a}$
does not change and $a$ does not receive impression $i$.
Therefore, we have
$p_{a,j}^{t+1} \gets p_{a,j}^{t}$ and
$D_{a,j}^{t+1} \gets D_{a,j}^{t}$ for $j \in \{0,1,2\}$.
Now we focus on updating these values for $a_1$ and $a_2$.

\paragraph{Case 1 ($|B| \le 1$ and $|B' \cup C| = 1$).}
First consider the case where $a_2$ does not exist.
The state $\Mark_{a_1}$ does not change,
so set $p_{a_1,j}^{t+1} \gets p_{a_1,j}^{t}$ for $j \in \{0,1,2\}$.
Since $i$ is assigned to $a_1$, we update
$D_{a_1,j}^{t+1} \gets \max(w_{i,a_1}, D_{a_1,j}^t)$ for each $j \in \{0,1,2\}$.
Here, the $\max$ operator transfers all probability mass from entries less than
$w_{i,a_1}$ to the value $w_{i,a_1}$.

\paragraph{Case 2 ($|B| \le 1$ and $|B' \cup C| \ge 2$).}
Now suppose $|B| \le 1$ and that there are two candidates $a_1$ and $a_2$.
The states $\Mark_{a_1}$ and $\Mark_{a_2}$ remain unchanged,
so set $p_{a_1,j}^{t+1} \gets p_{a_1,j}^{t}$
and $p_{a_2,j}^{t+1} \gets p_{a_2,j}^{t}$ for $j \in \{0,1,2\}$.
Impression~$i$ is assigned to $a_1$ or $a_2$ with equal probability $1/2$,
independent of their current states.
Therefore, we update the conditional distributions to be
$D_{a_1,j}^{t+1} \gets \frac{1}{2} \max(w_{i,a_1}, D_{a_1,j}^{t}) + \frac{1}{2} D_{a_1,j}^t$
and
$D_{a_2,j}^{t+1} \gets \frac{1}{2} D_{a_2,j}^t +  \frac{1}{2} \max(w_{i,a_2}, D_{a_2,j}^{t})$,
for each $j \in \{0,1,2\}$.

\paragraph{Case 3 ($|B| \ge 2$ and $\AdaptiveGain_{i,a_k} = 0$).}
Now suppose that $|B| \ge 2$ and that $\AdaptiveGain_{i,a_k} = 0$.
Recall that $\AdaptiveGain_{i,a_k}$ is a deterministic quantity governed
solely by the instance and arrival order.
The algorithm is guaranteed to enter the conditional statement on
line~25, so we update the priority probabilities to be
$p_{a_1,0}^{t+1} \gets 1, p_{a_1,1}^{t+1} \gets 0, p_{a_1,2}^{t+1} \gets 0$
and
$p_{a_2,0}^{t+1} \gets 1, p_{a_2,1}^{t+1} \gets 0, p_{a_2,2}^{t+1} \gets 0$
since the algorithm sets $\Mark_{a_1} \gets 0$ and $\Mark_{a_2} \gets 0$ on line~27.
The impression $i$ is randomly assigned to $a_1$ or $a_2$ with equal probability
on line~26, so we update the two well-defined conditional distributions to be
\begin{align*}
  D_{a_1,0}^{t+1} &\gets \frac{1}{2}\sum_{j=0}^2 p^t_{a_1,j} \max\left(w_{i,a_1}, D_{a_1,j}^{t}\right)
 + \frac{1}{2} \sum_{j=0}^2 p^t_{a_1,j}  D_{a_1,j}^{t}, \\
  D_{a_2,0}^{t+1} &\gets 
  \frac{1}{2} \sum_{j=0}^2 p^t_{a_2,j}  D_{a_2,j}^{t}
  + \frac{1}{2}\sum_{j=0}^2 p^t_{a_2,j} \max\left(w_{i,a_2}, D_{a_2,j}^{t}\right).
\end{align*}
The values of the other distributions
$D_{a_1,1}^{t+1}, D_{a_1,2}^{t+1}, D_{a_2,1}^{t+1}, D_{a_2,2}^{t+1}$
do not matter, so we leave them unchanged.
We note that the equations above can be simplified using the distributive 
property of the $\max$ operator, but the recurrences are easier to verify
in their current form.

\paragraph{Case 4 ($|B| \ge 2$ and $\AdaptiveGain_{i,a_k} > 0$).}
Now assume $|B| \ge 2$ and $\AdaptiveGain_{i,a_{k}} > 0$, where
$a_k$ is defined in line~21 of \Cref{alg:stochastic_greedy}.
In this case, the randomness of $\rv{R}_i$ and the current states of~$a_1$ and~$a_2$ determine how $i$ is assigned.
First, observe that we should set
$p_{a_1,j}^{t+1} \gets \frac{1}{3}$ and
$p_{a_2,j}^{t+1} \gets \frac{1}{3}$
for $j \in \{0,1,2\}$ since each of the three main branches is equally likely.
Now let us focus on computing the conditional distributions $D_{a_k,j}^{t+1}$
for $j \in \{0,1,2\}$.
If we condition on the value of $\Mark_{a_k}$ at the beginning of time $t+1$,
then we can determine how the algorithm branched at time $t$ based on the value
of $\rv{R}_i$.
This provides us with recurrence relations for the distributions $D_{a_k,j}^{t+1}$,
where we consider all possible previous states of $\Mark_{a_k}$ in each equation:
\begin{align*}
  D_{a_k,0}^{t+1} &\gets p_{a_k,2}^t \max\left(w_{i,a_k}, D_{a_k,2}^t\right)
                      + p_{a_k,1}^t D_{a_1,1}^t
                      + p_{a_k,0}^t \left(\frac{1}{2} \max\left(w_{i,a_k}, D_{a_k, 0}^t\right) + \frac{1}{2} D_{a_k, 0}^t\right),\\
  D_{a_k,1}^{t+1} &\gets p_{a_k,2}^t \max\left(w_{i,a_k} D_{a_k,2}^t\right) + p_{a_k,1}^t \max\left(w_{i,a_k} D_{a_k,1}^t\right) + p_{a_k,0}^t \max\left(w_{i,a_k} D_{a_k,0}^t\right),\\
  D_{a_k,2}^{t+1} &\gets p_{a_k,2}^t D_{a_k,2}^t + p_{a_k,1}^t D_{a_k,1}^t + p_{a_k,0}^t D_{a_k,0}^t.
\end{align*}
We can compute the distributions $D_{a_{3-k}, j}^{t+1}$ for $j \in \{0,1,2\}$ similarly, though in the adaptive
decisions we need to account for the probabilities $p_{a_{k},j'}^t$.
We use the equality $D_{a}^t = \sum_{j=0}^2 p_{a,j}^t D_{a,j}^t$ in the following
recurrences:
\begin{align*}
  D_{a_{3-k},0}^{t+1} &\gets
    p_{a_k,2}^t D_{a_{3-k}}^t
    +
    p_{a_k,1}^t \sum_{j=0}^2 p_{a_{3-k},j}^t \max\left(w_{i,a_{3-k}}, D_{a_{3-k},j}^t\right)
    +
    p_{a_k,0}^t \left(\frac{1}{2} D_{a_{3-k}}^t 
        + \frac{1}{2} \sum_{j=0}^2 p_{a_{3-k},j}^t \max\left(w_{i,a_{3-k}}, D_{a_{3-k},j}^t\right)  \right),\\
  D_{a_{3-k},1}^{t+1} &\gets
    p_{a_{3-k},2}^t \max\left(w_{i,a_{3-k}}, D_{a_{3-k},2}^{t}\right)
    + p_{a_{3-k},1}^t \max\left(w_{i,a_{3-k}}, D_{a_{3-k},1}^{t}\right)
    + p_{a_{3-k},0}^t \max\left(w_{i,a_{3-k}}, D_{a_{3-k},0}^{t}\right),\\
  D_{a_{3-k},2}^{t+1} &\gets
    p_{a_{3-k},2}^t D_{a_{3-k},2}^{t}
    + p_{a_{3-k},1}^t D_{a_{3-k},1}^{t}
    + p_{a_{3-k},0}^t D_{a_{3-k},0}^{t}.
\end{align*}

\paragraph{}
In all four cases, we have shown how to compute the probabilities $p_{a,j}^{t+1}$
and conditional distributions $D_{a,j}^{t+1}$
for each $a \in A$ and $j \in \{0,1,2\}$.
Therefore, by induction, we can maintain the distribution $D_{a}^t$ for $\MaxW_{a}^t$
at each time step $t$, and hence compute the exact values of $\E[\Gain_{i,a}]$.
\end{proof}

\section{Missing Analysis from \Cref{sec:analysis}}
\label{app:analysis}


Before we give the proofs of \Cref{lem:marginal_gain_lower_bound}
and \Cref{lem:adaptive_gain_upper_bound},
we present two self-contained, prerequisite lemmas.
For any random variable $\rv{X}$ and event $C$, define
$\E[\rv{X} : C]$ to be $\E[\rv{X} \mid C] \Pr(C)$ where
$\E[\rv{X} \mid C]$ is the expected value of $\rv{X}$ conditioned on $C$.
If this conditional probability is not well-defined,
we assume that $\E[\rv{X} : C] = 0$.
We use the following properties of the operators $(x)^+$ and $\E[\rv{X}:C]$
throughout this section.

\begin{restatable}[]{lemma}{PlusOperatorProperties}
\label{lem:plus_operator_properties}
For any three real numbers $u,v,w \in \R$, we have:
\begin{enumerate}
  \item $(w - \max\{u,v\})^+ \ge \parens{w-u}^+ - \parens{\max\{u,v\}-u}$,
  \item $(w-u)^+ \ge (w-\max\{v,u\})^+ + (v-u)^+ - (v-w)^+$.
\end{enumerate}
\end{restatable}

\begin{proof}
There are $3! = 6$ possible orderings for $u$, $v$, and $w$. 
We consider each case separately:

\begin{itemize}
\item\textbf{Case 1 ($w \ge u \ge v$).}
Property 1 is equivalent to $w - u \ge (w - u) - 0$, which is true.
Property 2 is equivalent to $w-u \ge (w-u) + 0 - 0$, which is also true.

\item\textbf{Case 2 ($w \ge v \ge u$).}
Property 1 is equivalent to $w - v \ge (w-u) - (v-u)$, which is true.
Property 2 is equivalent to $w-u \ge (w-v) + (v-u) - 0$, which is also true.

\item\textbf{Case 3 ($u \ge w \ge v$).}
Property 1 is equivalent to $0 \ge 0 - (u-u)$, which is true.
Property 2 is equivalent to $0 \ge 0 + 0 - 0$, which is also true.

\item\textbf{Case 4 ($u \ge v \ge w$).}
Property 1 is equivalent to $0 \ge 0 - (u-u)$, which is true.
Property 2 is equivalent to $0 \ge 0 + 0 - (v-w)$, which is also true.

\item\textbf{Case 5 ($v \ge w \ge u$).}
Property 1 is equivalent to $0 \ge (w-u) - (v-u)$, which is true.
Property 2 is equivalent to $w-u \ge 0 + (v-u) - (v-w)$, which is also true.

\item\textbf{Case 6 ($v \ge u \ge w$).}
Property 1 is equivalent to $0 \ge 0 - (v-u)$, which is true.
Property 2 is equivalent to $0 \ge 0 + (v-u) - (v-w)$, which is also true.
\end{itemize}
This completes the proof.
\end{proof}

\begin{restatable}[]{lemma}{DisjointEvents}
\label{lem:disjoint_events}
For any $k$ disjoint events $C_1,C_2,\dots,C_k$ and nonnegative random variable
$\rv{X}$, we have
  \[
    \E[\rv{X}] \ge \sum_{i=1}^k \E[\rv{X}:C_i].
  \]
If these disjoint events span the probability space, that is, 
$\sum_{i=1}^k \Pr\parens{C_i} = 1$, the inequality can be replaced by equality,
even if $\rv{X}$ is not nonnegative.
\end{restatable}

\begin{proof}
Let $\overline{C}$ be the event that none of $C_1,C_2,\dots,C_k$
occur. The $k+1$ events $\overline{C},C_1,C_2,\dots,C_k$ are all disjoint and
span the entire probability space.
Therefore, for any random variable $\rv{X}$ (not necessarily nonnegative),
the law of total expectation gives us
\[
  \E\bracks*{\rv{X}} = \E\bracks*{\rv{X} \mid \overline{C}} \Pr\parens*{\overline{C}}
  + \sum_{i=1}^k \E\bracks*{\rv{X} \mid C_i} \Pr\parens*{C_i}
= \E\bracks*{\rv{X} : \overline{C}} + \sum_{i=1}^k \E\bracks*{\rv{X} : C_i},
\]
which proves the second part of the claim since $\Pr(\overline{C}) = 0$ if
$C_1,C_2,\dots,C_k$ span the probability space.
For the first part, if $\rv{X}$ is nonnegative then
$\E[\rv{X} : \overline{C}]$ is also nonnegative, and therefore
$\E\bracks*{\rv{X}} \ge \sum_{i=1}^k \E\bracks{\rv{X} : C_i}$,
which concludes the proof.
\end{proof}

\subsection{Proof of \Cref{lem:marginal_gain_lower_bound}}

\MarginalGainLowerBound*

\begin{proof}
Recall that the variables
$\AdaptiveGain_{i,a}$ and $S_a$ are not random variables as show in
\Cref{lem:deterministic}---they are completely determined by the arrival
order of the impressions.
Recall also that $k \in \{1,2\}$ is set in line~21 of \Cref{alg:stochastic_greedy} such that
$\AdaptiveGain_{i,a_{k}} \ge \AdaptiveGain_{i,a_{3-k}}$.
If $\AdaptiveGain_{i,a_{k}}=0$, then we also have
$\AdaptiveGain_{i,a_{3-k}}=0$, so we need to show that
\[
  \E[\MarginalGain_i] \ge \frac{\E[\Gain_{i,a_1}] + \E[\Gain_{i,a_2}]}{2}.
\]
This is evident because $i$ is assigned to $a_1$ or $a_2$ with equal
probability on line~26.
Therefore, we focus on the case $\AdaptiveGain_{i,a_{k}} > 0$ for the rest of the proof.

The $\AdaptiveGain_{i,a}$ variables are computed in terms of the expected values
$\E[\Gain_{i,a}]$ and therefore do not depend on the state of the algorithm
(i.e., independent of the previous coin tosses $\rv{R}_i$) by
\Cref{lem:expected_gain_values}.
Therefore, conditioning on the event $\AdaptiveGain_{i,a_k} > 0$
does not affect the distribution of $\rv{R}_i$ variables.

Impression $i$ is assigned to $a_1$ or $a_2$ with equal probability
(and also independently to prior decisions of the algorithm)
unless $\rv{R}_i \in [0,\sfrac{1}{3})$
and $\Mark_{a_k} \in \{1,2\}$.
This is the only cases where $i$ is not assigned to~$a_1$ or $a_2$ symmetrically.
We also note that $\Mark_{a_k} \in \{1,2\}$ is associated with the
events $\rv{R}_{i'} \in [\sfrac{1}{3},\sfrac{2}{3})$
or $\rv{R}_{i'} \in [\sfrac{2}{3},1)$,
where $i'$ is equal to $\Index(a_k)$ at the beginning of time $t=t_{i}$
when $i$ arrives.
Let $t' = t_{i'} < t$ be the time that $i'$ arrives.
Observe that the random variables $\rv{R}_i$ and $\rv{R}_{i'}$ are independent of each other.
Therefore, by the law of total expectation,
the expected marginal gain $\E[\MarginalGain_i]$ is equal to
\[
  \frac{\E\bracks*{\Gain_{i,a_1}} + \E\bracks*{\Gain_{i,a_2}}}{3}
  +
  \frac{\E\bracks*{\Gain_{i,a_{k}} : C_2} + \E\bracks*{\Gain_{i,a_{3-k}} : C_1}}{3}
  +
  \frac{\E\bracks*{\Gain_{i,a_{k}} : C_0} + \E\bracks*{\Gain_{i,a_{3-k}} : C_0}}{6},
\]
where $C_2$, $C_1$, and $C_0$ are the events that $\Mark_{a_k}$ is equal to
$2$, $1$, and $0$, respectively.
These events partition the space, 
so \Cref{lem:disjoint_events} implies 
$(\E[\Gain_{i,a_{k}} : C_2] + \E[\Gain_{i,a_{k}} : C_0])/6 =
 (\E[\Gain_{i,a_{k}}] - \E[\Gain_{i,a_{k}} : C_1])/6$.
An analogous decomposition also holds for $\E[\Gain_{i,a_{3-k}}]/6$.
Therefore, $\E[\MarginalGain_i]$ is equal to
\[
  \frac{\E\bracks*{\Gain_{i,a_1}} + \E\bracks*{\Gain_{i,a_2}}}{2}
  +
  \frac{\E\bracks*{\Gain_{i,a_{k}} : C_2} + \E\bracks*{\Gain_{i,a_{3-k}} : C_1}}{6}
  -
  \frac{\E\bracks*{\Gain_{i,a_{k}} : C_1} + \E\bracks*{\Gain_{i,a_{3-k}} : C_2}}{6}.
\]
Since $\AdaptiveGain_{i,a_k} \ge \AdaptiveGain_{i,a_{3-k}}$,
it suffices to prove the inequalities
\begin{align}
\label{eqn:desired_inequality_one}
  \frac{\E[\Gain_{i,a_k} : C_2] - \E[\Gain_{i,a_k} : C_1]}{6} \ge 2\cdot \AdaptiveGain_{i,a_k}
\end{align}
and
\begin{align}
  &\E[\Gain_{i,a_{3-k}} : C_1] - \E[\Gain_{i,a_{3-k}} : C_2] \ge 0. \label{eqn:desired_inequality_two}
\end{align}
to complete the proof.

\paragraph{Proof of Inequality \Cref{eqn:desired_inequality_one}.}
Conditioning on the event $C_2$, we can apply Property~1 in \Cref{lem:plus_operator_properties}
to get
\[
  \Gain_{i,a_{k}} =
  \parens*{w_{i,a_k} - \MaxW_{a_k}^{t-1}}^+
  \ge \parens*{w_{i,a_k} - \MaxW_{a_k}^{t'-1}}^+
    - \parens*{\MaxW_{a_k}^{t-1} - \MaxW_{a_k}^{t'}}.
\]
Recall that $t' = t_{\Index(a_k)}$ and
note that no impression is assigned to $a_k$ at time $t'$ according to the event $C_2$.
Therefore, we have $\MaxW_{a_k}^{t'-1} = \MaxW_{a_k}^{t'}$.
Taking the conditional expectation implies that
\begin{align}
\label{eqn:C2_event}
  \E\bracks*{\Gain_{i,a_k} : C_2}
  &\ge \E\bracks*{\parens*{w_{i,a_k} - \MaxW_{a_k}^{t'-1}}^+ : C_2}
     - \E\bracks*{\MaxW_{a_k}^{t-1} - \MaxW_{a_k}^{t'} : C_2} \notag\\
  &\ge \frac{\E\bracks*{\parens*{w_{i,a_k} - \MaxW_{a_k}^{t'-1}}^+}}{3}
     - \E\bracks*{\MaxW_{a_k}^{t-1} - \MaxW_{a_k}^{t'}},
\end{align}
where the second inequality holds for the following reasons.
The random variable $\MaxW_{a_k}^{t'-1}$ is independent of event $C_2$,
so
$\E\bracks{(w_{i,a_k} - \MaxW_{a_k}^{t'-1})^+ : C_2}
 = \E\bracks{(w_{i,a_k} - \MaxW_{a_k}^{t'-1})^+}/3$.
We can also apply \Cref{lem:disjoint_events} for the nonnegative term
$\MaxW_{a_k}^{t-1} - \MaxW_{a_k}^{t'}$ and to get
$\E[ \MaxW_{a_k}^{t-1} - \MaxW_{a_k}^{t'} : C_2] \le \E[\MaxW_{a_k}^{t-1} - \MaxW_{a_k}^{t'}]$.

The term $\MaxW_{a_k}^{t-1} - \MaxW_{a_k}^{t'}$ represents all assignments
to $a_k$ in the time range $[t'+1,t-1]$, inclusive.
In this interval, $\Index(a_k)$ has remained the same since it was last
set to $i'$ at time $t'$.
Therefore, any impression $i''$ assigned to $a_k$ in this time period has been
allocated in \CaseTwo of the algorithm (i.e., the case $|B| \le 1$).
Since we increment $S_{a_k}$ for each of these assignments accordingly
(by either $M_{i''}$ or $M_{i''}/2$ depending on whether $a_k$ is the only candidate or not),
the expected increase of $\MaxW_{a_k}$ in this time period
is upper bounded by $S_{a_k}$.
Formally, we have $\E[\MaxW_{a_k}^{t-1} - \MaxW_{a_k}^{t'}] \le S_{a_k}$.
Therefore, it follows from \Cref{eqn:C2_event} that
\begin{align}
\label{eqn:C2_event_part2}
  \E\bracks*{\Gain_{i,a_k} : C_2} \ge
  \frac{\E\bracks*{\parens*{w_{i,a_{k}} - \MaxW_{a_k}^{t'-1}}^+}}{3} - S_{a_k}.
\end{align}

By Property 2 of \Cref{lem:plus_operator_properties}, we always have the inequality
\begin{align}
\label{eqn:C2_event_part3}
  \parens*{w_{i,a_k} - \MaxW_{a_k}^{t'-1}}^+
  \ge
  \parens*{w_{i,a_k} - \max\set*{w_{i',a_k}, \MaxW_{a_k}^{t'-1}}}^+
    + \parens*{w_{i',a_k} - \MaxW_{a_k}^{t'-1}}^+
    - \parens*{w_{i',a_k} - w_{i,a_k}}^+.
\end{align}
By combining inequalities
\Cref{eqn:C2_event_part2} and \Cref{eqn:C2_event_part3}, it follows that
\begin{align}
\label{eqn:gain_conditioned_on_C2}
  \E[\Gain_{i,a_k} \hspace{-0.08cm}:\hspace{-0.08cm} C_2]
  &\ge \frac{\E\bracks*{\parens*{w_{i,a_k} \hspace{-0.07cm}- \max\set*{w_{i',a_k}, \MaxW_{a_k}^{t'-1}}}^+}}{3}
    +
    \frac{\E\bracks*{\parens*{w_{i',a_k} \hspace{-0.07cm}- \MaxW_{a_k}^{t'-1}}^+}}{3}
    -
    \frac{\parens*{w_{i',a_k} \hspace{-0.07cm}- w_{i,a_k}}^+}{3} - S_{a_k} \notag\\
  &\ge \E\bracks*{\Gain_{i,a_k} : C_1} + 
    \frac{\E\bracks*{\Gain_{i',a_k}}}{3} - \frac{\parens*{w_{i',a_k} - w_{i,a_k}}^+}{3} - S_{a_k},
\end{align}
where \Cref{eqn:gain_conditioned_on_C2} is proved as follows.
Conditioning on event $C_1$, we know that impression $i'$ was assigned to $a_k$.
Thus, $\Gain_{i,a_k} = \parens{w_{i,a_k} - \MaxW_{a_k}^{t-1}}^+$
is at most $\parens{w_{i,a_k} - \max\set{w_{i',a_k}, \MaxW_{a_k}^{t'-1}}}^+$.
Observing that $\MaxW_{a_k}^{t'-1}$ is independent of $C_1$, we have
\[
  \frac{\E\bracks*{\parens*{w_{i,a_k} - \max\set*{w_{i',a_k}, \MaxW_{a_k}^{t'-1}}}^+}}{3} \ge \E\bracks*{\Gain_{i,a_k} : C_1},
\]
which proves \Cref{eqn:gain_conditioned_on_C2}.
Note that the only assumption we used to prove \Cref{eqn:gain_conditioned_on_C2}
is that $\AdaptiveGain_{i,a_k}$ is positive.
Therefore, the definition of $\AdaptiveGain_{i,a_k}$ in line 8
of \Cref{alg:stochastic_greedy} and inequality \Cref{eqn:gain_conditioned_on_C2}
imply that
\[
  \E[\Gain_{i,a_k} : C_2] - \E[\Gain_{i,a_k} : C_1] \ge 12 \cdot \AdaptiveGain_{i,a_k},
\]
which completes the proof of inequality~\Cref{eqn:desired_inequality_one}.

\paragraph{Proof of Inequality \Cref{eqn:desired_inequality_two}.}
Let $a'$ be $\Partner(a_k)$ at time $t-1$ (right before $a_{3-k}$ becomes
the partner of~$a_k$). Impression $i'$ is assigned to either $a_k$ or $a'$.
The $\Index$ values of these two advertisers are unchanged in the interval
$[t'+1,t-1]$, otherwise $\Color(a_k)$ would be $\green$ at the
beginning of time~$t$, which contradicts the positivity of
$\AdaptiveGain_{i,a_k}$.
Moreover, the advertiser $a' \ne 0$ is well-defined by the positivity of
$\AdaptiveGain_{i,a_k}$ since $\Color_a(a_k) = \blue$.
Therefore, all impressions assigned to $a_k$ or~$a'$
in this interval must be
non-adaptive decisions (i.e., are assigned in \CaseTwo). These impressions
are assigned independent to the value that $\rv{R}_{i'}$ takes for impression $i'$.
Each impression that arrives before time~$t$ and has neither $a_k$ nor $a'$
as its candidates is also assigned independent to the value of $\rv{R}_{i'}$.
In particular, conditioning on the events $\Mark_{a_k} = 1$
or $\Mark_{a_k} = 2$ does not change the distribution of $\MaxW_{a}^{t-1}$
for any $a \in A \setminus \{a_k, a'\}$.
Therefore, if $a_{3-k}$ is not the same as $a'$, then 
$\E[\Gain_{i,a_{3-k}} : C_1] = \E[\Gain_{i,a_{3-k}} : C_2]$,
which proves the inequality.
If $a_{3-k} = a'$, then all assignments to $a_{3-k}$ at times
$T=\{1,2,\dots,t-1\}\setminus\{t'\}$ are independent of both events
$\Mark_{a_k} = 1$ and $\Mark_{a_k} = 2$.
If we fix the set of impressions assigned to $a_{3-k}$ at times in $T$,
conditioning on $\Mark_{a_k}=1$ compared to conditioning on $\Mark_{a_k}=2$
can only decrease $\MaxW^{t-1}_{a_{3-k}}$, which further implies 
$\E[\Gain_{i,a_{3-k}} : C_1] \ge \E[\Gain_{i,a_{3-k}} : C_2]$.
This completes the proof of inequality \Cref{eqn:desired_inequality_two},
and therefore the proof of the lemma.
\end{proof}

\subsection{Proof of \Cref{lem:adaptive_gain_upper_bound}}

\AdaptiveGainUpperBound*

\begin{proof}
If $\AdaptiveGain_{i,a} = 0$, then the claim is trivial.
Therefore, assume that
\[
  \AdaptiveGain_{i,a} = \frac{\E\bracks*{\Gain_{i',a}}/3 
     - \parens*{w_{i',a} - w_{i,a}}^{+}/3 - S_a}{12} > 0,
\]
where $i'$ is equal to $\Index(a)$ at the beginning of time $t=t_i$.
Indices are set on line~17 of \Cref{alg:stochastic_greedy}, so
$\AdaptiveGain_{i,a} > 0$ implies that $i'$ is assigned to an advertiser in \CaseOne.
Without loss of generality, assume~$a$ is the first choice of $i'$
(i.e., $a = a_1$ and $a \ne a_2$).
The proof for the other case is identical by swapping the $\Mark_{a}=1$
and $\Mark_{a}=2$ terms.

We focus on the events where $i'$ is assigned in lines 29--30 of
\Cref{alg:stochastic_greedy}, which correspond to setting $\Mark_{a} = 1$ or $\Mark_{a} = 2$.
Let $C_1$ be the event that $i'$ is assigned to $a$ in line~29,
or, equivalently, the event $\Mark_a = 1$.
We note that $i'$ might be assigned to $a$ in \CaseOne but this is not
included in the event $C_1$.
Similarly, let $C_2$ be the event that $i'$ is assigned to the second candidate
in line~30. Let $t'=t_{i'}$ be the time $i'$ arrives.
Conditioning on $C_2$, we can apply \Cref{eqn:gain_conditioned_on_C2} from
the proof of \Cref{lem:marginal_gain_lower_bound}, which gives us
\begin{align}
  \label{eqn:gain_lower_bound_again}
  \E\bracks*{\Gain_{i,a} : C_2} &\ge \E\bracks*{\Gain_{i,a} : C_1}
    + \frac{\E\bracks*{\Gain_{i',a}}}{3}
    - \frac{\parens*{w_{i',a} - w_{i,a}}^+}{3}
    - S_a \notag\\
  &\ge
    \frac{\E\bracks*{\Gain_{i',a}}}{3}
    - \frac{\parens*{w_{i',a} - w_{i,a}}^+}{3}
    - S_a.
\end{align}
Since $\AdaptiveGain_{i,a} > 0$,
inequality \Cref{eqn:gain_lower_bound_again} implies 
$\E[\Gain_{i,a} : C_2] \ge 12 \cdot \AdaptiveGain_{i,a}$.
Therefore, we have
$\E[\Gain_{i,a}] \ge \E[\Gain_{i,a} : C_2] \ge 12 \cdot \AdaptiveGain_{i,a}$
by \Cref{lem:disjoint_events}
since $\Gain_{i,a}$ random variables are always nonnegative.
\end{proof}

\end{document}